\documentclass[aps,prb,superscriptaddress,twocolumn,tightenlines,showpacs,floatfix,amsmath,amssymb,pdftex]{revtex4-1}
\usepackage{graphicx}
\usepackage{xcolor}
\usepackage{physics}
\usepackage{dsfont}
\usepackage{hyperref}
\usepackage{chngcntr}

\begin{document}

\title{The inconsistency of linear dynamics and Born's rule}

\author{Lotte Mertens}
\affiliation{Institute for Theoretical Physics Amsterdam,
University of Amsterdam, Science Park 904, 1098 XH Amsterdam, The Netherlands}
\affiliation{Institute for Theoretical Solid State Physics, IFW Dresden, Helmholtzstr. 20, 01069 Dresden, Germany}
\author{Matthijs Wesseling}
\author{Niels Vercauteren}
\author{Alonso Corrales-Salazar}
\author{Jasper van Wezel}
\affiliation{Institute for Theoretical Physics Amsterdam,
University of Amsterdam, Science Park 904, 1098 XH Amsterdam, The Netherlands}

\date{\today}

\begin{abstract}
Modern experiments using nanoscale devices come ever closer to bridging the divide between the quantum and classical realms, bringing experimental tests of objective collapse theories that propose alterations to Schr\"{o}dinger's equation within reach. Such objective collapse theories aim to explain the emergence of classical dynamics in the thermodynamic limit and hence resolve the inconsistency that exists within the axioms of quantum mechanics. Here, we show that requiring the emergence of Born's rule for relative frequencies of measurement outcomes without imposing them as part of any axiom, implies that such objective collapse theories cannot be linear. Previous suggestions for a proof of the emergence of Born's rule in classes of problems that include linear objective collapse theories are analysed and shown to include hidden assumptions.
\end{abstract}

\maketitle

\section{Introduction}
While quantum mechanics is the best-tested theory within physics to date~\cite{peskin_schroeder_2020}, there is a well-known inconsistency in its axioms~\cite{komar_1962, wigner_1963, whynotdec, penrosegrav}, 
known as the `measurement problem'. From the moment the framework of quantum mechanics was first formulated, this problem has fuelled a search for the connection between the experimentally verified probabilistic outcomes of measurement and the fundamentally deterministic time evolution prescribed by the Schr\"odinger equation~\cite{manyworlds, bohm, copenhagen,CSL,GRW, Mastereq}.
This quest for understanding how non-deterministic measurement arises from deterministic quantum dynamics continues unabated as an active field of research today~\cite{overview, Penrose, symmetry, schlosshauer_kofler_zeilinger_2013}. Approaches to addressing the measurement problem can be divided into two categories, either attempting to give alternative interpretations of the mathematical structures featured in the quantum theory, or attempting to supplement or alter the laws of quantum physics to include the emergence of measurement at macroscopic scales~\cite{overview}. The latter are known as objective collapse theories. 

Several experimental techniques have recently been developed to explore the region between the microscopic realm, where quantum dynamics has been verified to proceed according to the Schr\"odinger equation to extremely high accuracy, and the macroscopic realm, where measurement devices yield probabilistic outcomes for quantum measurements~\cite{superposmirror, underground, vinante_mezzena_falferi_carlesso_bassi_2017, carlesso_bassi_falferi_vinante_2016, wit_welker_heeck_buters_eerkens_koning_meer_bouwmeester_oosterkamp_2019, arndt_nairz_vos-andreae_keller_zouw_zeilinger_1999, mooij_nazarov_2006, andrews_1997, christian_2005}. 

It is in this unexplored region intermediate between microscopic and macroscopic superpositions that objective collapse theories predict the quantum-classical crossover to take place~\cite{CSL, Penrose, Mastereq}, 
and yield observable differences in their physical predictions from interpretation-based approaches. The direct observation of the the mesoscopic realm thus necessitates a theoretical exploration of both the dynamics predicted by different classes of objective collapse theories, and the postulates underlying their predictions in this regime. Like the bounds imposed by experimental observation, consistency requirements on the theoretical postulates and dynamics may then be used to classify and constrain objective collapse theories. 

In this article, we classify objective collapse theories according to the requirement that relative frequencies associated with measurement outcomes, known as Born's rule, emerge without imposing them as part of any axiom~\cite{BR,MWBR,ZehBR}. We show that imposing this physical constraint rules out theories for quantum measurement based on either linear or unitary generators of time evolution. In particular, we study the dynamics of mesoscopic two-state systems imposed by a generic time evolution operator and determine its late-time behaviour. Demanding that individual solutions should be stable and that collectively they obey Born's rule, leads to a set of constraints that cannot be satisfied in any linear or unitary theories. We also formulate a minimal non-linear objective collapse model for the two-state system that does reproduce Born's rule without assuming it at any point.

\section{Born's rule}
\label{sec:Copen}
Regardless of interpretation, the measurement of a quantum state is commonly accepted to be separable into two stages~\cite{wigner_1963, neumann_beyer_wigner_hofstadter_1955, zurek_1981}. In the first instance, a microscopic object is entangled with a measurement machine. For a two-state superposition, this can be written as:
\begin{align}
    &\left(\alpha \ket{0} + \beta \ket{1} \,\right) \otimes \ket{M_{\text{init}}} \notag \\ 
    \to~ &\alpha \ket{0} \otimes \ket{M_0} + \beta \ket{1} \otimes \ket{M_1}.
    \label{eq:initstate}
\end{align}
Here, the states $\ket{0}$ and $\ket{1}$ are two distinct quantum states of the microscopic object, while $\ket{M_{\text{init}}}$ is the initial state of the measurement machine, and $\ket{M_0}$ and $\ket{M_1}$ are states of the measurement machine that have a macroscopic object (the `pointer') indicating measurement outcomes $0$ and $1$ respectively. That is, the states $\ket{M_0}$ and $\ket{M_1}$ form the `pointer basis'. This initial stage of the measurement process can be realised using unitary quantum dynamics according to the Schr\"odinger equation and may for the sake of simplicity be assumed to be infinitely fast~\cite{anglin_paz_zurek_1997}. 

All interpretations and objective collapse theories agree up to this point in the measurement process~\cite{wigner_1963}. What they disagree on is how, given the state in Eq.~\eqref{eq:initstate}, an observer registers one, and only one, outcome on the measurement machine.

So-called interpretations of quantum mechanics posit that the entangled state of Eq.~\eqref{eq:initstate} lives on forever and that the reason why observers only see one outcome lies in the physical interpretation of what the entangled wave function represents. These include the splitting of realities~\cite{manyworlds}, the separation of the wave function into a physical state and pilot waves~\cite{bohm}, and others~\cite{overview}.

Objective collapse theories on the other hand, introduce a dynamical process that reduces the entangled state of Eq.~\eqref{eq:initstate} to just one randomly selected pointer state in each measurement, indicating only one of the possible measurement outcomes. These theories necessarily involve an addition to or modification of the Schr\"odinger equation. Well-known examples include continuous spontaneous localisation (CSL) theories \cite{pearle, CSL}, the Ghirardi–Rimini–Weber (GRW) model \cite{GRW}, and  mechanisms related to the influence of gravity on quantum dynamics~\cite{Penrose, diosi_1987, Wezel_Brink}. In all of these objective collapse theories, the dynamics involved in the second stage of measurement takes a finite, non-zero time to complete, and this collapse time depends on the size, the mass, or some other property of the measurement machine. This way, microscopic objects are guaranteed to be impervious to the modifications imposed on the Schr\"odinger equations for any measurable time, while the dynamics of macroscopic pointers will be so dominated by its effect that collapse occurs almost instantaneously~\cite{overview}.  

Mesoscopic experiments currently being developed \cite{underground, carlesso_bassi_falferi_vinante_2016, arndt_nairz_vos-andreae_keller_zouw_zeilinger_1999, mooij_nazarov_2006, andrews_1997, christian_2005} may probe the dynamics of objects that are heavy or large enough to feel modifications to the Schr\"odinger equation, but light or small enough for the ensuing dynamics to take a measurably long time to complete. They include for example a mirror in an optical interferometer~\cite{superposmirror}, a low temperature mechanical resonator \cite{wit_welker_heeck_buters_eerkens_koning_meer_bouwmeester_oosterkamp_2019} or free falling masses in space \cite{vinante_mezzena_falferi_carlesso_bassi_2017}.

All these experiments directly target the transition between quantum and classical physics by investigating whether the dynamics starting from the state in Eq.~\eqref{eq:initstate} deviates from that predicted by the Schr\"odinger equation. Since alternative interpretations of quantum mechanics adhere to the Schr\"odinger equation at all scales, they predict these types of experiments to yield no result. In the remainder of this article, we will therefore focus exclusively on objective collapse theories and the measurable dynamical processes predicted by them.

In the limit of the measurement machine being very heavy or large, all objective collapse theories must reproduce our everyday experience of quantum measurement. This implies that these theories possess at least the following three characteristics:
\begin{enumerate}
    \item Preferred basis: an initial superposition such as that of Eq.~\eqref{eq:initstate} is dynamically reduced to a single state within a pointer basis.
    \item Stability: a macroscopic measurement machine in a single pointer state should not spontaneously evolve out of that state at any observable timescale.
    \item Born's rule: the relative frequency with which a particular measurement outcome results from the process of quantum measurement should equal the squared weight of the corresponding pointer state in the initial superposition.
\end{enumerate}
The first characteristic formalises the observation that macroscopic measurement machines indicate only a single measurement outcome after each experiment, while the second prevents the registered outcome of a measurement from changing after the measurement process has completed. The final characteristic is commonly known as Born's rule and has been experimentally verified for macroscopic measurement machines to great accuracy~\cite{sinha_couteau_jennewein_laflamme_weihs_2010, pleinert_von}. Notice that its formulation here assumes the initial superposition to be normalised. 

Since the aim of objective collapse theories is to provide a complete description of the measurement process, all three characteristics should emerge from the state evolution during measurement. That is, one should be able to derive them from the predicted collapse dynamics itself even if one does not know about their existence beforehand~\cite{Zurek}. In some theories for quantum measurement the characteristic that they give rise to Born's rule is built into the theory as an axiomatic assumption governing either additions to the Schr\"odinger equation~\cite{GRW,CSL}, or the initial state of the universe~\cite{bohm}. Other theories, however, have been suggested to give rise to Born's rule without assuming it in any way~\cite{Zurek, symmetry, valentini_westman_2005}. 

In particular, a rigorous way of constructing objective collapse theories that are guaranteed not to assume or depend on Born's rule, is to only consider alterations to the Schr\"odinger equation that add linear operators to its time evolution generator. The elements of linear operators in any matrix representation are independent of the wave function that the operator acts on and can therefore not contain any information related to Born's rule. Moreover, they typically fall into a class of theories which have been suggested to necessarily give rise to the emergence of Born's rule~\cite{Zurek,vanwezelprb}. The arguments underlying this suggestion depend only on the structure of entanglement between a quantum state and its environment and are independent of the precise dynamics during measurement. They were introduced in the context of decoherence~\cite{Zurek}, and have been applied in several well-known approaches to the quantum measurement problem that rely on decoherence for establishing a pointer basis and Born's rule~\cite{Deutsch_99,Wallace_03,Wallace_12,Valentini,Saunders}.

\section{Dynamics on the Bloch Sphere}
Any objective collapse theory will have to be able to at least describe the measurement dynamics of a two-state superposition, like that of Eq.~\eqref{eq:initstate}. Focusing on that simplest possible situation, we will consider the time evolution of a general two-state system parameterized on the Bloch sphere:
\begin{align}
\label{eq:initialwave}
    \ket{\psi_0} = n e^{i \chi} \left [ e^{i\frac{\phi}{2}}\cos{(\theta/2)}\ket{0}+e^{-i\frac{\phi}{2}}\sin{(\theta/2)}\ket{1}\right].
\end{align}
Here, the states $\ket{0}$ and $\ket{1}$ represent the products of microscopic and pointer states in Eq.~\eqref{eq:initstate}. Since the dynamics during measurement is dominated by the dynamics of the measurement machine, one can equivalently think of the states $\ket{0}$ and $\ket{1}$ as just the pointer states of the measurement device itself. The amplitudes of the coefficients are determined by the angle $\theta \in [0, \pi]$, while their relative phase is given by $\phi \in [0, 2\pi]$. The norm $n$ and overall phase $\chi$ are shown in Appendix~\ref{AppA} to not influence the time dependence of the amplitudes and relative phase, even for general (not necessarily unitary) time evolution operators. The relative weights and phases of the two states in the superposition can thus be represented by a point on the Bloch sphere.

\begin{figure}[tb]
  \centering
    \includegraphics[width=\columnwidth]{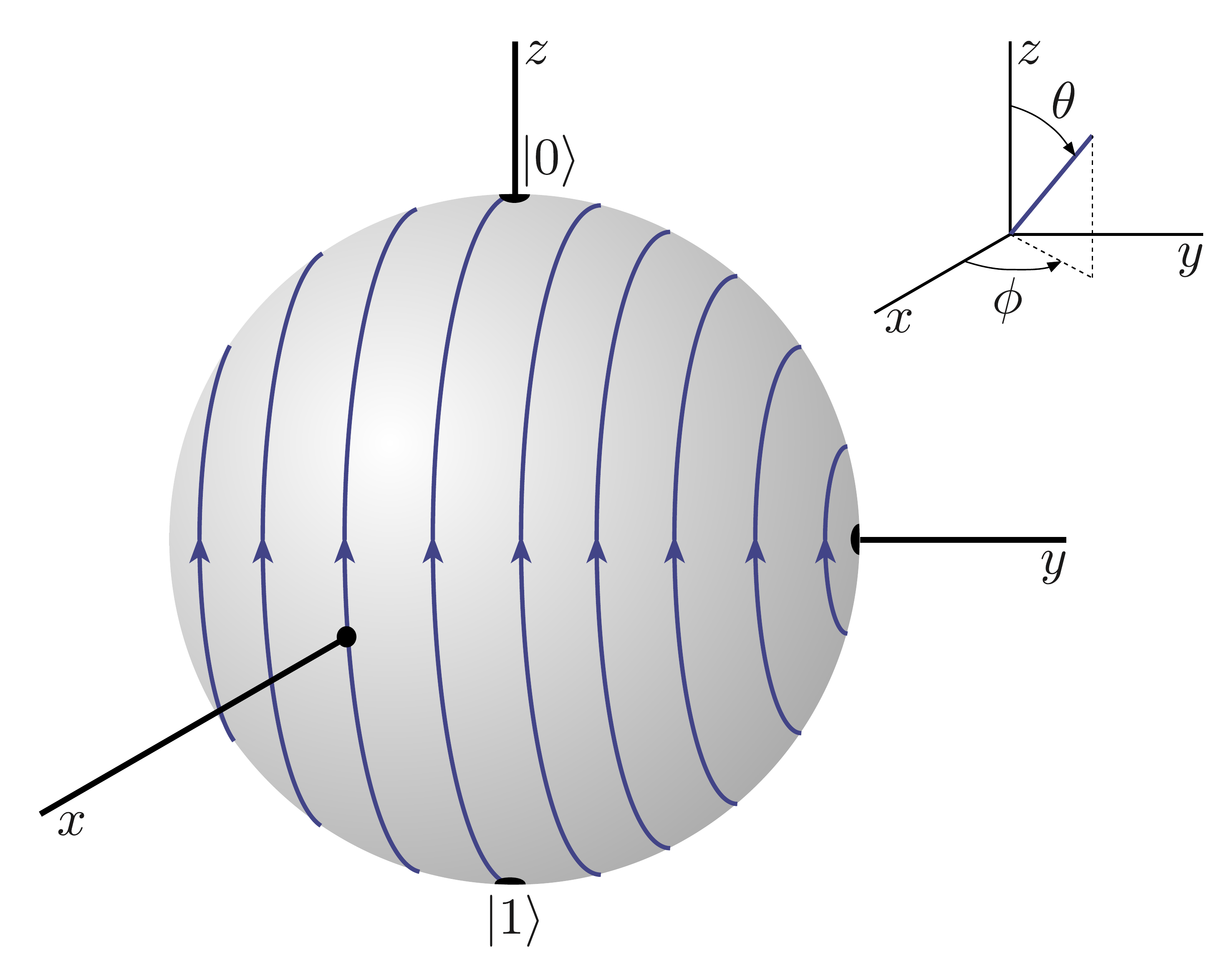}
    \caption{Unitary evolution on the Bloch sphere. Each point on the Bloch sphere represents a different state with angles $\theta$ and $\phi$. The flow lines indicate Rabi oscillations generated by a Hamiltonian proportional to the Pauli matrix $\hat{\sigma}_y$. Different lines represent oscillations with different initial values for $\theta$ and $\phi$.}
    \label{fig:rabi2}
\end{figure}

Evolution of the wave function can be visualised on the Bloch sphere by flow lines. Under unitary time evolution, generated by the usual quantum mechanical time evolution operator $e^{-i \hat{H} dt/\hbar}$ with $\hat{H}$ the Hamiltonian, the flow on the Bloch sphere is conservative and consists of closed cycles known as Rabi oscillations, as shown in Fig.~\ref{fig:rabi2}. More generally, we can describe time evolution in any linear theory for two-state systems as being generated by the operator $e^{-i \hat{G} dt} $, with $\hat{G}$ a general $2\times 2$ matrix. As long as the generator of time evolution is unitary (but not necessarily linear), the flow on the Bloch sphere is conservative and the flow lines are all closed loops. That is, each initial state will undergo indefinite periodic time evolution.

An objective collapse theory, on the other hand, should cause an initial state of the form of Eq.~\eqref{eq:initialwave} to eventually end up in either one of the pointer states (characteristic 1). Moreover, upon reaching a pointer state, the system should cease to evolve (characteristic 2). On the Bloch sphere, this means that the state needs to end up at either the north or south pole, and that the pole towards which it evolves needs to be a stable end-point of the evolution. Stable end points of flow lines are either attractive fixed points or limit cycles, but as limit cycles are inherently nonlinear, we consider only fixed points here~\cite{chaos}. 

Different measurements starting from the same initial state should evolve to state $\ket{0}$ with probability $\cos^2(\theta/2)$ and to state $\ket{1}$ with probability $\sin^2(\theta/2)$ (characteristic 3). The flow lines on the Bloch sphere can therefore not be fixed entirely by just the initial state. Rather, for any given initial state, there must be a set of possible evolutions, one of which is randomly selected each measurement. We take an agnostic approach to the physics or degrees of freedom that control the random selection and introduce a single (non-local) random variable $\lambda$ that determines the particular set of flow lines selected in any individual measurement. Born's rule then emerges if the relative frequency with which flow lines terminating at a particular pointer state are selected, equals its squared weight in the initial superposition. Notice that the distribution of values that $\lambda$ can take must be independent of the (initial) state of the system being measured, in order to avoid introducing Born's rule in the definition of the time evolution generator.

\section{Stable collapse to a pointer state}
The general time evolution generated by $e^{-i\hat{G}dt}$, with $\hat{G}$ a linear operator, can be represented in terms of a $2\times2$ matrix with 8 real parameters: 
\begin{align}
\label{eq:G}
    \hat{G} = \begin{pmatrix} \ket{0} & \ket{1} \end{pmatrix}
    \begin{pmatrix}
\alpha_r+i\alpha_i & \beta_r+i\beta_i\\
\gamma_r + i\gamma_i & \delta_r + i\delta_i 
\end{pmatrix}
\begin{pmatrix} \bra{0} \\ \bra{1} \end{pmatrix}.
\end{align}
The generator $\hat{G}$ can be written as the sum of a Hermitian and an anti-Hermitian contribution. Depending on the coupling constant or energy scale governing the strength of the anti-Hermitian part, the time evolution of microscopic systems will be practically unaffected by it on any observable time scale, while the dynamics of macroscopic systems are instantly dominated by the anti-Hermitian contribution~\cite{pearle, GRW, symmetry, Wezel_Brink, diosi_1987}. 

In order for $\hat{G}$ to give rise to measurement dynamics, resulting in a stable final state at either the north or south pole of the Bloch sphere, the flow lines it generates must have an attractive fixed point on at least one of the poles. These flow lines can be found explicitly by constructing the time derivatives of the parameters in the state of Eq.~\eqref{eq:initialwave}, as shown in Appendix~\ref{AppA}. To find possible fixed points of the flow, however, it is more instructive to directly consider the equation $\partial_t \ket{\psi(t)}=-i\hat{G} \ket{\psi(t)}$. Since $\hat{G}$ is a linear operator working within the two-state Hilbert space spanned by $\ket{0}$ and $\ket{1}$, it will have two eigenstates, and the general solution of the time evolution equation can be written as~\cite{chaos}:
\begin{align}
\ket{\psi(t)} = e^{-i\lambda_1 t} C_1 \ket{\psi_1} + e^{-i\lambda_2 t} C_2 \ket{\psi_2}.
\label{eq:psit}
\end{align}
Here, $\ket{\psi_{1,2}}$ are the eigenstates of $\hat{G}$ and $\lambda_{1,2}$ the corresponding eigenvalues. The coefficients are given by $C_j=\braket{\psi_j}{\psi(0)}$. For the moment, we assumed that the generator $\hat{G}$ does not depend on time. 

From equation~\eqref{eq:psit} it is immediately clear that if both eigenvalues are real, the dynamics does not have any fixed points. This corresponds to the case of a purely Hermitian $\hat{G}$ and unitary time evolution. If either one or both of the eigenvalues have an imaginary component however, the relative weight of one of the eigenstates will grow exponentially with time.
Notice that in this process, the total norm of the wave function is not conserved. This is consistent with the fact that we do not \emph{a priori} interpret the squared norm as a probability for finding particular measurement outcomes. Rather, if at late times the state of the measurement machine is guaranteed to always consist of only a single pointer state (either $\ket{\psi_1}$ or $\ket{\psi_2}$), that state can be taken to be the outcome of the measurement process, regardless of its norm. In that case, Born's rule is equivalent to writing the expectation value for a physical quantity $O$ as:
\begin{align}
\bar{O} = \frac{\bra{\psi} \hat{O} \ket{\psi}}{\braket{\psi}{\psi}}.
\end{align}
Here, $\bar{O}$ is the expectation value of the observable represented by the (Hermitian) operator $\hat{O}$. This re-definition of the axiom relating physically observed expectation values to a mathematical property  of operators and states does not affect any of the predictions of standard, unitary quantum mechanics.

\begin{figure}[tb]
  \centering
    \includegraphics[width=\columnwidth]{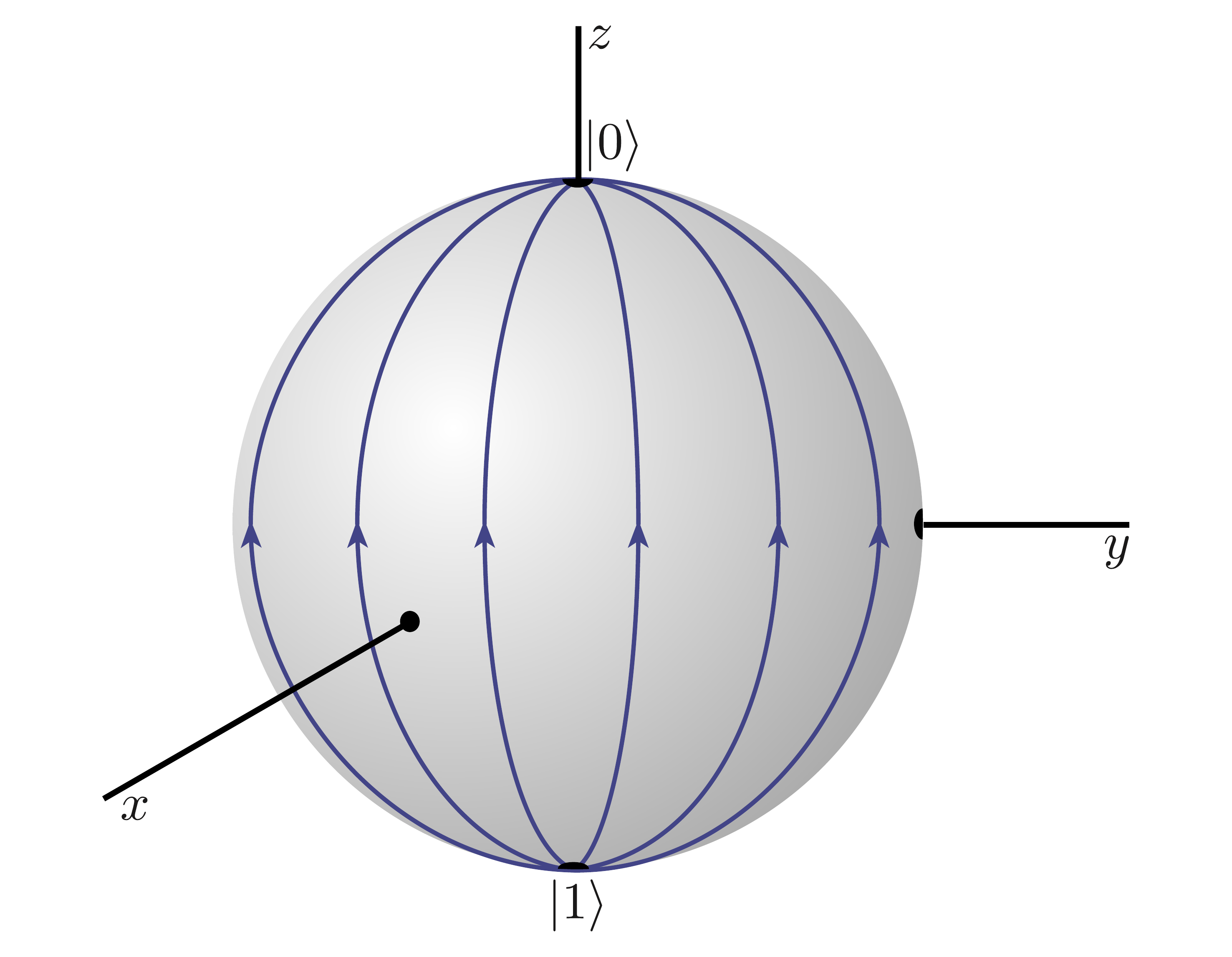}
    \caption{Schematic representation of the flow generated by a purely diagonal non-Hermitian generator of time evolution. The state $\ket{0}$ is an attractive fixed point of the flow, while $\ket{1}$ is a repulsive fixed point.}
    \label{fig:source}
\end{figure}

Because $\hat{G}$ acts within a two-state Hilbert space, equation~\eqref{eq:psit} in general has two fixed points. The absence of limit cycles for linear flow then guarantees that one will be a source of flow lines and the other a sink~\cite{chaos}, as shown in figure~\ref{fig:source}. The only two exceptions possible are purely unitary quantum dynamics, in which the fixed points become centres of rotation, and the exceptional situation in which the two fixed points coalesce into a single half-attractive, half-repulsive point. Even this latter case, however, still has all flow lines terminating in the single fixed point.

The first characteristic of measurement dynamics, that an initial superposition of pointer states should evolve to just a single pointer state, is satisfied if the attractive fixed point created by $\hat{G}$ is a pointer state. In fact, if there is a separation of time scales between the dynamics induced by the Hermitian and anti-Hermitian parts of $\hat{G}$ when acting on macroscopic objects~\cite{diosi_1987,symmetry,GRW,CSL,Penrose},
only the attractive fixed point of the anti-Hermitian part by itself needs to be a pointer state. For our present model, this implies that the two pointer states $\ket{0}$ and $\ket{1}$ should be the two orthogonal eigenstates of the anti-Hermitian part of $\hat{G}$. That is, the anti-Hermitian part is diagonal, so that $\beta_r=\gamma_r$ and $\beta_i=-\gamma_i$. The state $\ket{0}$ is then an attractive fixed point of the non-unitary flow if $\alpha_i > \delta_i$, and $\ket{1}$ is attractive for $\alpha_i < \delta_i$. The second characteristic, that the final states in the evolution are stable, is automatically satisfied as long as $\hat{G}$ is time-independent, since no flow lines escape from the attractive fixed point (see figure~\ref{fig:source}).

\section{Born's rule}
For objective collapse to satisfy the third characteristic of measurement dynamics, that it yields Born's rule, it must necessarily contain a stochastic component. This can be introduced into the description in terms of a general generator $\hat{G}$ by having its components depend on one or more parameters that are randomly drawn from a given probability distribution. These parameters could have a physical interpretation in, for example, the existence of some fundamental field beyond quantum mechanics that fluctuates in time and influences the dynamics of measurement machines~\cite{symmetry}. Within the evolution generated by $\hat{G}$, the random parameters will then determine which of the pointer states corresponds to an attractive fixed point of the dynamics.

For macroscopic measurement machines, $\hat{G}$ is expected to generate collapse dynamics that is almost instantaneous. The randomly fluctuating field will then be effectively static within the time it takes for the collapse to complete and only a single value for each random parameter needs to be considered for any individual measurement process. Depending on how the random parameters influence the sign of $\alpha_i-\delta_i$, either the state $\ket{0}$ or the state $\ket{1}$ is then selected to be the sole attractive fixed point of the flow on the Bloch sphere. That pointer state thus becomes the measurement outcome regardless of any property of the initial state. In particular, this means the measurement outcomes cannot adhere to Born's rule, which prescribes a distribution of outcomes that depends on the state being measured.

The selection of which pointer state is the attractive fixed point can not depend on the initial state, since a linear operator $\hat{G}$ is by definition independent of the state it acts on. This implies in particular that also the distribution of random variables appearing in a linear generator does not depend on the state it acts on. 

Relaxing the constraint on the random variables and considering objective collapse dynamics for which $\hat{G}$ is non-linear only through the distribution of its stochastic components, is possible~\cite{GRW, CSL}. However, Born's rule then typically does not emerge from the dynamics, but is rather hard-wired into the dependence of random variables on the state to be measured. The axiom of expectation values adhering to Born's rule is then replaced by the axiom of random variables adhering to a distribution that results in Born's rule. Here, we avoid such axioms altogether and instead focus on linear stochastic processes only. As noted before, however, the linearity of $\hat{G}$ and its independence of the initial state imply that it cannot generate nearly instantaneous collapse dynamics consistent with the emergence of Born's rule.

That leaves the possibility of the random parameters fluctuating faster than the typical time it takes for measurement to complete. This is especially relevant for practical measurement machines, which may be large compared to the quantum particles whose properties they measure, but which are nevertheless finite in size and mass. The short but finite time scale associated with that large size may well be longer than the typical time it takes for a randomly fluctuating parameter to significantly change its value.

In terms of the two-state evolution induced by $\hat{G}$, evolving values for the random parameters should cause the sign of $\alpha_i-\delta_i$ to randomly change in time. This could in principle yield probabilities for measurement outcomes that depend on the initial state, since not all points on the Bloch sphere travel equally far to the attractive fixed point within the time that the random variable has an approximately fixed value. The fluctuating dynamics, however, pose a different problem, as fluctuations of the fixed points from being attractive to repulsive and back again, causes the evolution to lose its stable end points. 

That a fixed point cannot be reached even in the infinite time limit, is clear from the fact that for every value $\alpha_i-\delta_i=a$, there is the value $-a$ with precisely reversed flow lines. Both values must occur with equal likelihood owing to the fact that the random parameters cannot have any preference for either of the two possible outcomes. For any initial state, the likelihood of a random variable occurring that causes a flow towards one pole of the Bloch sphere is therefore equal to the likelihood of flowing towards the other.

\subsection{Fuzzy collapse}
A possible way around the lack of precise stable end states could be the concept of fuzzy collapse~\cite{Fuzzy}. That is, the speed of flow across the Bloch sphere might in principle be such that once a state is within some cut-off $\delta\theta$ of the poles, it takes a time longer than any realistic human time scale for the state to leave that region. We can then effectively consider the evolution as being stopped when the fuzzy region surrounding any pole has been reached, making it possible to assign a definite measurement outcome to all states within the fuzzy region surrounding the poles rather than just the poles themselves. 

To calculate the probability of reaching $\theta = \delta\theta$ before reaching $\theta=\pi-\delta\theta$, we can map the evolution on the Bloch sphere onto a one dimensional random walk. Each evolution line in the flow diagram can be mapped onto a straight one-dimensional line. Depending on the sign of $\alpha_i-\delta_i$ the state will move either up or down the line. The size of the step taken along the line in a given time interval is determined by the values of the parameters in $\hat{G}$. The average step size going up, however, must be equal to the average step size going down from any given state, owing to the fact that the distribution of random values may not imply a preference for either of the pointer states. Assuming that many steps are necessary to reach the fuzzy collapse region and thus taking the limit of infinitesimal step size, then yields the probability of reaching one particular end point without having reached the other before~\cite{firstpassage, Lotte}: 

\begin{align}
\label{eq:Pvary}
    P \left( \theta_0  \to 0+\delta\theta \right) = \frac{1}{2} + \frac{1}{2}\frac{\ln\left[\cot(\theta_0/2)\right]}{\ln\left[\cot(\delta \theta/2)\right]}.
\end{align}
Here, $\theta_0$ is the initial value of the parameter $\theta$ in the initial state of equation~\eqref{eq:initialwave}. The probability $P(\theta_0 \to \delta \theta)$ is called the splitting probability~\cite{firstpassage}. In the limit $\delta\theta\to 0$, where the fuzzy collapse region contains just the pointer states, the splitting probability becomes flat and independent of the initial state. For non-zero values of $\delta\theta$, the splitting probabilities do depend on $\theta_0$, but they never reproduce Born's rule.

\subsection{Numerical simulation of collapse dynamics}
\begin{figure}[tb]
 \centering 
     \includegraphics[width=\columnwidth]{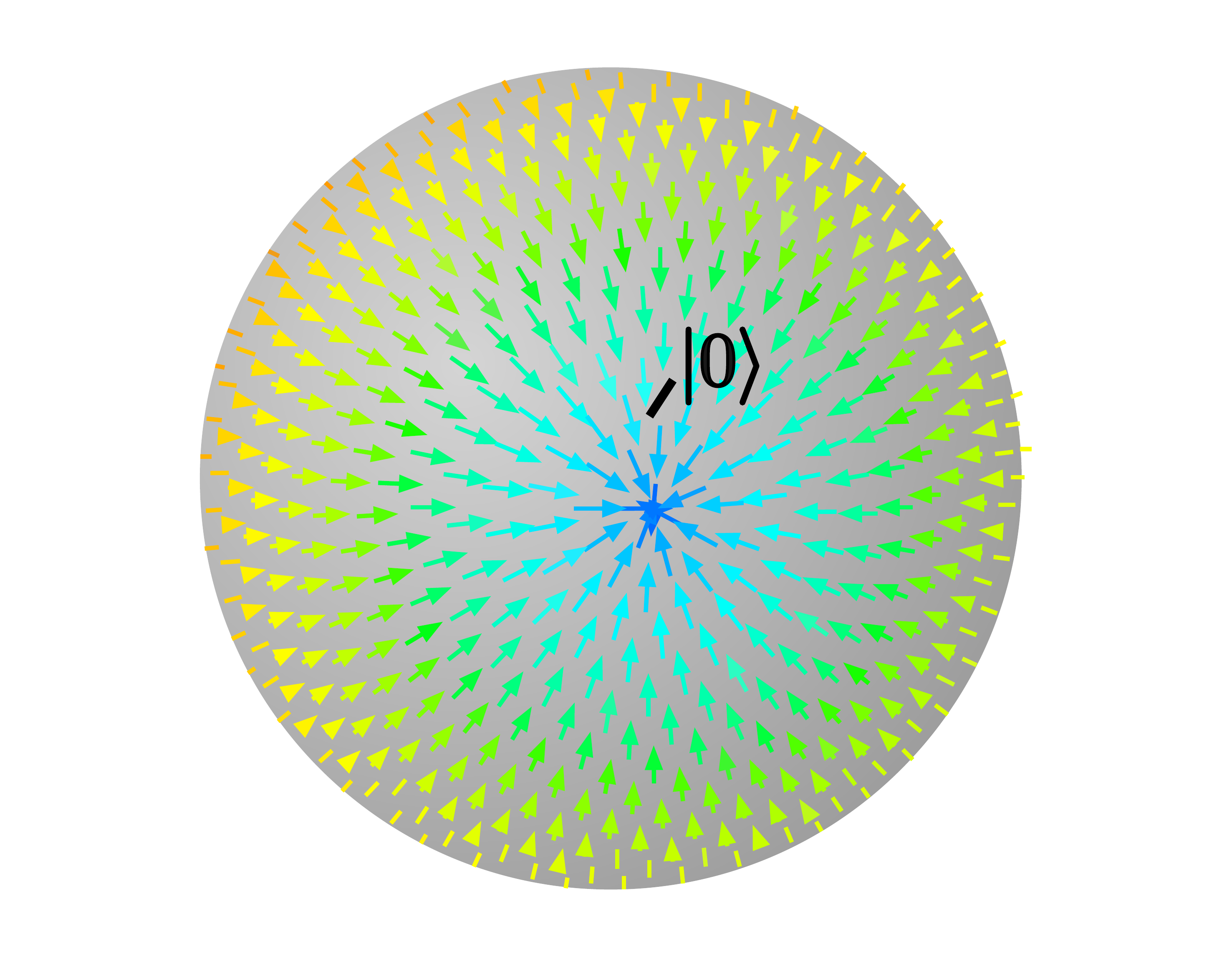}
     \caption{The velocity field $(\dot{\theta},\dot{\phi})$ of the flow on the Bloch sphere generated by $\hat{G}$ using the (arbitrarily chosen) parameter values $\alpha_i-\delta_i=1$, $\beta_i=0$, $\gamma_i=0.5$, $\beta_r=0.1$, and $\gamma_r=0$. 
     The orientations of the arrows represent the direction of the local velocity, while their colours indicate the local speed of the flow, ranging from blue (lowest) to red (highest). The speed decreases to zero at the fixed point. }
     \label{fig:shift}
 \end{figure}

For non-zero step size, the evolution induced by the generator $\hat{G}$ of equation~\eqref{eq:G} can be numerically simulated. As shown in Appendix~\ref{AppA}, a Taylor expansion of the equation $\partial_t \ket{\psi(t)}=-i\hat{G}\ket{\psi(t)}$ directly yields the time dependence of the Bloch sphere coordinates $\theta(t)$ and $\phi(t)$. For time-independent parameters, the velocity $(\dot{\theta},\dot{\phi})$ can be plotted directly on the Bloch sphere to visualize the flow lines and fixed points of the dynamics, as shown in figure~\ref{fig:shift}. For dynamically fluctuating parameters, the velocities can be numerically integrated to yield the dynamics starting from any point on the Bloch sphere.

\begin{figure}[tb]
    \centering
    \includegraphics[width=\columnwidth]{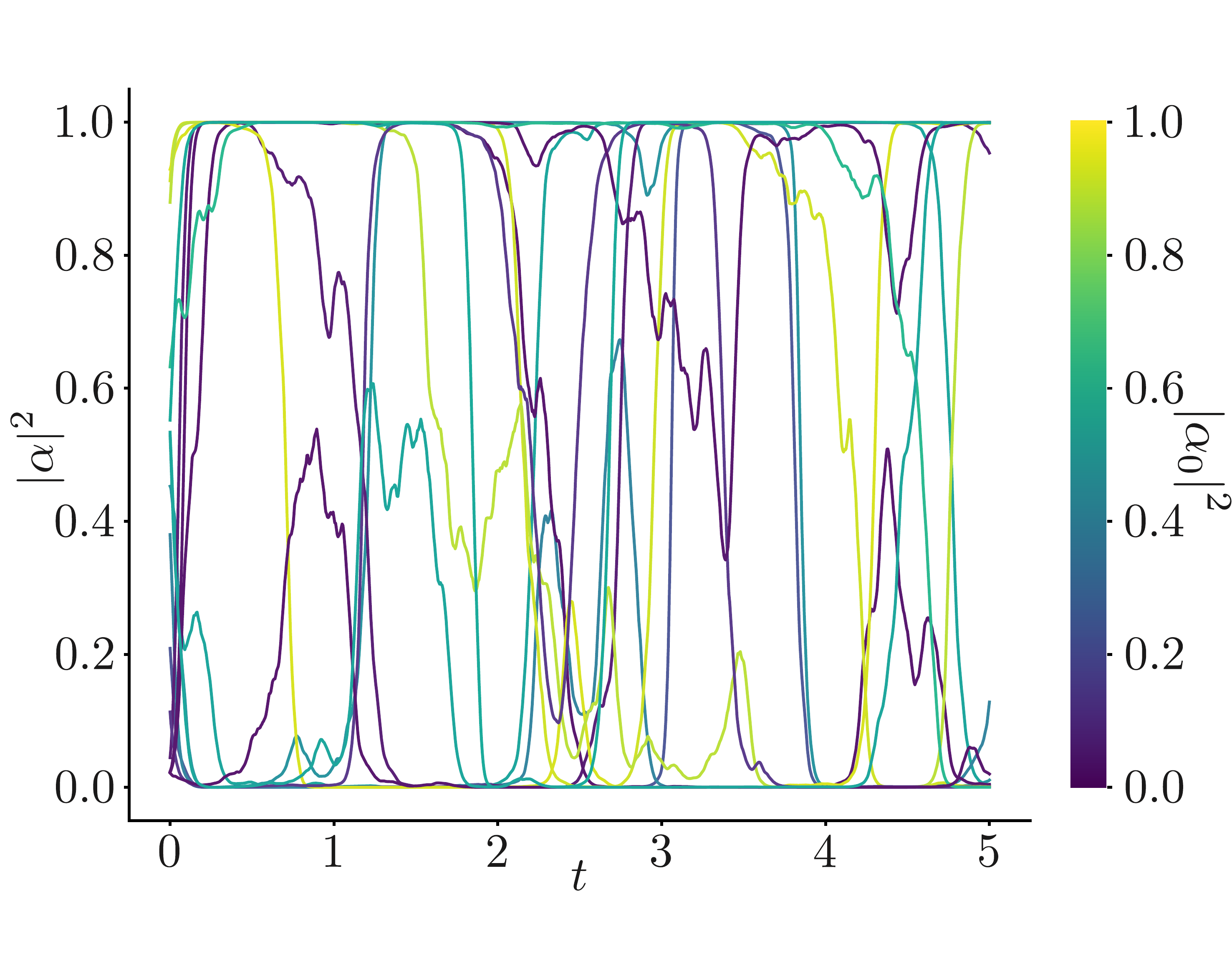}
    \caption{Instability of the collapse dynamics in the presence Each time step $dt=0.005$, the value of $\alpha_i$ is drawn from a Gaussian distribution with standard deviation $20$, while all other parameters in $G$ are zero. Initial states with different initial weights $|\alpha_0|^2 = |\cos(\theta_0/2)|^2$ are indicated by differently coloured lines. The absolute weight $|\alpha(t)|^2=|\cos(\theta(t)/2)|^2$ is plotted against time $t$ and shows that the state may get arbitrarily close to a fixed point, but cannot stay there indefinitely.}
    \label{fig:instability}
\end{figure}

Taking as an example the flow defined by having $\alpha_i$ randomly fluctuating in time and all other parameters in $G$ being zero, figure~\ref{fig:instability} shows the evolution of $|\alpha(t)|^2=\cos^2(\theta(t)/2)$ as a function of time. The random switches between $|\alpha|=0$ and $|\alpha|=1$ are clearly visible for all initial states and cannot be avoided for any choice of parameter values. It is straightforward to check that the typical time required to go from any initial value $\theta_0$ to any other value $\theta_1$ is equal to the typical time to return from $\theta_1$ to $\theta_0$.

\begin{figure}[tb]
    \centering
    \includegraphics[width=\columnwidth]{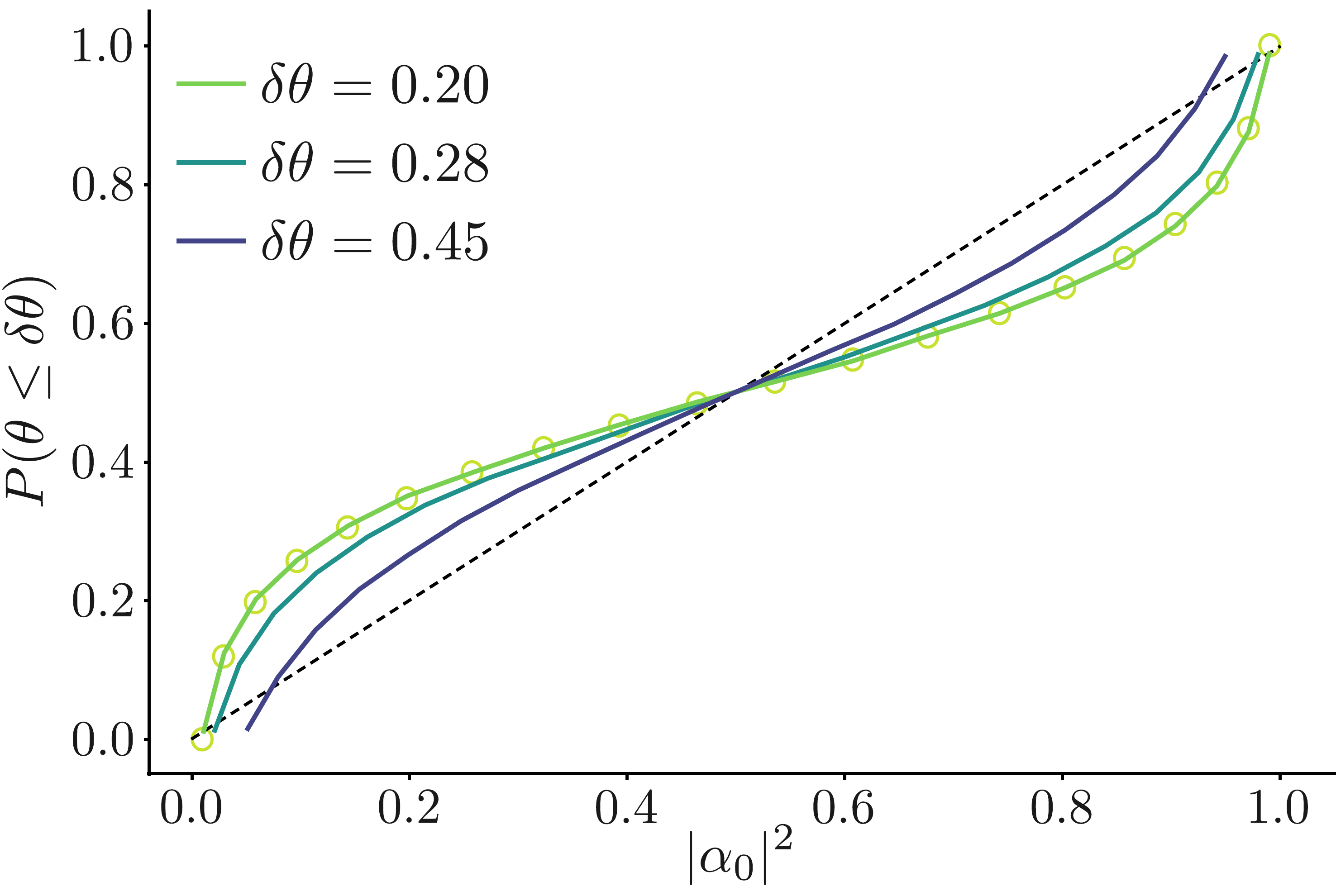}
    \caption{Fuzzy collapse statistics. The relative frequency of evolutions coming within $\delta \theta$ of the state $\ket{0}$ before coming within $\delta \theta$ of $\ket{1}$, plotted as a function of the initial initial weight  $|\alpha_0|^2=\cos^2(\theta_0/2)$. The dashed black line represents Born's rule, while the circles depict the splitting probability of equation~\ref{eq:Pvary} for $\delta\theta=0.20$. For each value of $\alpha_0$, $10^5$ instances of the dynamics are calculated for a maximum of $10^4$ time steps with $dt=0.05$. Each time step, the value of $\alpha_i-\delta_i$ is randomly chosen from a flat distribution in the interval $[-1, 1]$, while all other parameters in $G$ are zero.  
    }
    \label{fig:varystat}
\end{figure}

Stopping the time evolution as soon as the value of $\theta(t)$ comes within $\delta \theta$ of either zero or pi, the frequency of different fuzzy collapse outcomes can be simulated. The resulting statistics are shown in figure~\ref{fig:varystat}. They approach the splitting probabilities in the continuum limit of zero step size. In the limit of $\delta \theta$ going to zero, the observed frequencies become constant and independent of the initial state again.

It should be noted that the results illustrated here for specific values of the parameters appearing in $\hat{G}$ are in fact generic. For any set of parameters with fixed points at the poles of the Bloch sphere, including any contribution from random variables with an even distribution around $\alpha_i-\delta_i=0$, the dynamics does not have stable fixed points, and the statistics of even fuzzy measurement outcomes do not adhere to Born's rule.

\section{Born's rule from envariance}
The fact that linear models for objective collapse cannot give rise to Born's rule is surprising, given that they fall into a class of models in which the emergence of Born's rule has previously been suggested to be unavoidable~\cite{Zurek,vanwezelprb}. This suggestion was first made in the context of decoherence and used the possibility of quantum states entangling with an external environment~\cite{Zurek}. The assumptions that enter the suggested proof of Born's rule emerging, however, do not depend on the actual presence, influence, or dynamics of any environmental states. Essentially the same suggested proof has therefore also been applied in several other well-known approaches to the quantum measurement problem, including the pilot wave and many-worlds theories~\cite{Deutsch_99,Wallace_03,Wallace_12,Valentini,Saunders,vanwezelprb}.

The same assumptions also apply in the current framework of linear dynamics on a two-state superposition, and the failure of the current framework to give rise to Born's rule thus indicates the presence of additional hidden assumptions in the suggested proof. To be specific, the analysis of Ref.~\cite{Zurek} starts from the initial state of equation~\eqref{eq:initialwave}, and assumes that there is some process that will eventually reduce this initial state to one of two possible pointer states. This could be a trace over environmental degrees of freedom, as in the theory of decoherence, or non-unitary time evolution, as considered here. It is also assumed that the process leading to a final pointer state depends only on the weights occurring in the initial state superposition, and not on the states being superposed. In the non-unitary evolution, this is guaranteed by $\hat{G}$ being linear.

The first step in the suggested proof that the probability for ending up in a given pointer state adheres to Born's rule, is then to notice that in the special case of equation~\eqref{eq:initialwave} having equal weights for the two pointer states, it can be made `envariant'~\cite{Zurek}. That is, we could imagine entangling the two-state system with a second, external degree of freedom, so that the combined state becomes:
\begin{align}
    \ket{\psi} = \alpha \left( \ket{0}\ket{a} + \ket{1}\ket{b} \right).
\end{align}
Here, $\ket{0,1}$ denote the pointer states of the system, while $\ket{a,b}$ are environmental states. This combined state is envariant in the sense that the effect of a swap operation interchanging the system states can be undone by a swap operation on the environment~\cite{Zurek}. Assuming that the environmental degree of freedom is causally disconnected from the system degree of freedom, an action on the environment should not influence the statistics of any measurement outcomes on the system. This implies that a swap operation on the system should not influence the measurement outcomes, since it can be undone by a causally disconnected swap on the environment. The probability for the system to end up in state $\ket{0}$ must therefore be equal to the probability for the system to end up in state $\ket{1}$ (see Appendix~\ref{AppB} for details). 

Notice that this conclusion does not require the environmental state to actually exist or be present. It suffices that it could in principle exist and that any local measurements on the system should not be able to allow for conclusions about the existence of the environmental state to be made, because it is causally disconnected. That requirement is enough to force the probabilities for finding either pointer state to be equal. This is independent even of the physical process leading to the observation of only a single pointer state and therefore applies equally to objective collapse models and alternative interpretations of quantum mechanics.

The suggested proof for the emergence of Born's rule eventually extends the above reasoning to a state with unequal weight superpositions, which again may be entangled with a causally disconnected external degree of freedom (see Appendix~\ref{AppB}):
\begin{align}
    \ket{\psi} = \alpha \ket{0}\ket{a} + \beta \ket{1}\ket{b}.
\end{align}
The coefficients in this state may be assumed to be real without loss of generality and likewise, we may assume an external degree of freedom within an arbitrarily large Hilbert space. This allows us to choose basis in which to expand $\ket{a}$ and $\ket {b}$ in such a way that  the state $\ket{\psi}$ can be written as:
\begin{align}
    \ket{\psi} = \sqrt{\frac{1}{N}} \left[ \sum_{i=1}^n \ket{0}\ket{i} + \sum_{j=n+1}^{n+m} \ket{1}\ket{j} \right].
    \label{eq:zurekstate}
\end{align}
Because of the arbitrary size of the external Hilbert space, we can choose the rational numbers $n/N$ and $m/N$ such that they approximate $|\alpha|^2$ and $|\beta|^2$ with arbitrary precision.

The equal weights with which all states appear in equation~\eqref{eq:zurekstate} again allows for an argument based envariance to be made. That is, we can imagine there may exist a second environmental state, causally disconnected from both the system and the original environment, but entangled with both:
\begin{align}
    \ket{\psi} = \sqrt{\frac{1}{N}} \left[ \sum_{i=1}^n \ket{0}\ket{i}\ket{e_i} + \sum_{j=n+1}^{n+m} \ket{1}\ket{j}\ket{e_j} \right].
    \label{eq:zurekstate2}
\end{align}
This state is invariant in the sense that a swap operation between two states of the original environment can be undone by a swap operation between two states of the causally disconnected second environment. Using the same arguments as before (see also Appendix~\ref{AppB}), this implies that the probabilities for ending up in any one of the states $\ket{0}\ket{i}\ket{e_i}$ or $\ket{1}\ket{j}\ket{e_j}$ must all be equal.

The final step in the suggested proof is then to argue that because all states $\ket{0}\ket{i}\ket{e_i}$ contain the system state $\ket{0}$ and are orthogonal, the probability of ending up with the system in the state $\ket{0}$ is equal to $n$ times the probability for ending in one of the states $\ket{0}\ket{i}\ket{e_i}$. That is, the probability for ending up in $\ket{0}$ is suggested to equal $n/N$, in accordance with Born's rule.

As we showed, however, this final conclusion cannot be realised in any linear collapse model for two-state systems. The apparent paradox is resolved by a hidden assumption in the final step in the analysis based on envariance. The combined probability for ending up in any one of the states $\ket{0}\ket{i}\ket{e_i}$ is not the same as the probability for finding the single state $\sum_i \ket{0}\ket{i}\ket{e_i}$. The former implies that the result of the measurement is one of the states $\ket{0}\ket{i}\ket{e_i}$ (or a diagonal density matrix), whereas the latter corresponds to a pure state superposition of all of these states. The observation that all components in equation~\eqref{eq:zurekstate2} have equal probability of being the final state in a combined measurement (or decoherence process) of the system and the first environmental state, does not imply anything about the probabilities involved in a measurement (or decoherence process) registering the state of just the system.

\section{Minimal example for objective collapse}
\begin{figure}[tb]
    \centering
    \includegraphics[width=\columnwidth]{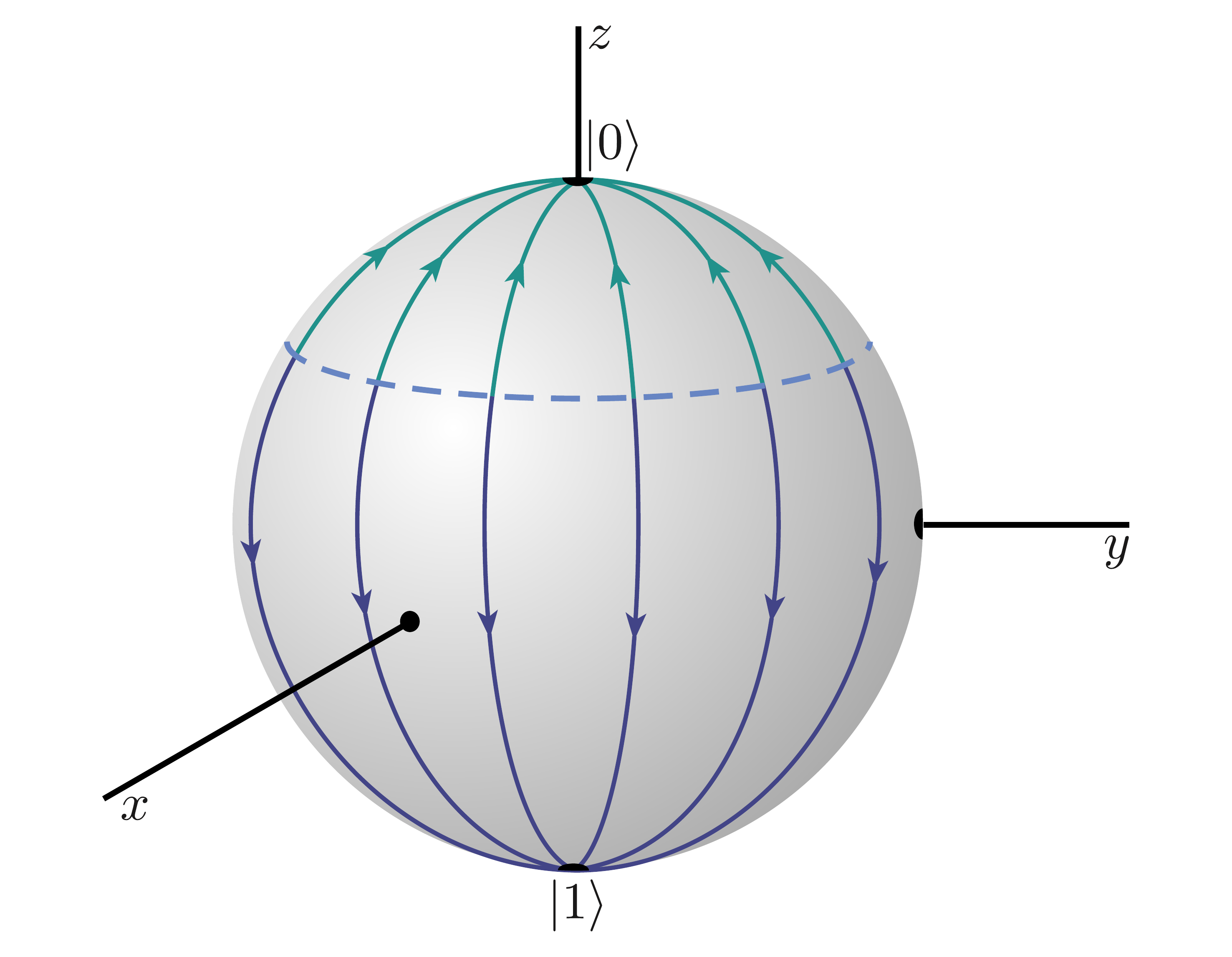}
    \caption{The flow lines for the minimal objective collapse model defined by equation~\eqref{eq:thet1}. The dashed blue line shows the seperatrix dividing areas of the Bloch sphere that flow towards distinct pointer states.}
    \label{fig:sinksourceline}
\end{figure}

The underlying reason that no linear model for objective collapse can yield Born's rule, is that the linearity forbids the initial state from having any influence on the dynamics. A non-linear model for objective collapse dynamics can also be written with a time evolution operator of the from $e^{-i\hat{G}dt}$, but in that case the matrix elements of $\hat{G}$ explicitly depend on the state it acts on. In terms of the flow it generates on the Bloch sphere, elements beyond simple sources and sinks of flow lines become allowed in the presence of a non-linear generator. With those, a pattern of flow lines for the two-state system that meets all three characteristics of quantum measurement can be constructed. A minimal example is defined by:
\begin{align}
\label{eq:thet1}
    \dot{\theta} &= \sin(\theta)\left(\lambda - \cos(\theta)  \right) \notag \\
    \dot{\phi} &= 0.
\end{align}
Here, $\lambda$ is a time-independent random variable with a flat distribution in the interval $[-1, 1]$. The flow lines generated by these equations are shown in figure~\ref{fig:sinksourceline}. They have two attractive fixed points, at the poles of the Bloch sphere. Their basins of attraction are bounded by the separatrix $\theta = \arccos(\lambda)$, indicated by a dashed blue line in figure~\ref{fig:sinksourceline}. Thus, if the initial state lies above the dashed blue line ($\theta_0 > \arccos(\lambda)$), it will flow towards $\ket{0}$ while if the initial state lies below the blue line ($\theta_0 < \arccos(\lambda)$) it will flow towards $\ket{1}$.

Since the values of $\theta_0$ and $\lambda$ fully determine the late time behaviour of the system, the probability of flowing to $\ket{0}$, given a distribution of random variables $f(\lambda)$, is given by:
\begin{align}
    P(\ket{0}) = \int_{-1}^1 f(\lambda) \Theta(\cos(\theta_0) -\lambda) d\lambda
\end{align}
Here, $\Theta$ is the Heaviside step function. For a flat distribution of $\lambda$ in the interval $[-1, 1]$ the probability becomes $1/2 \int_{-1}^{\cos{\theta_0}}d\lambda = cos(\theta_0) + 1 = \cos^2(\theta_0/2)$. In other words, the probability precisely matches Born's rule. 

Because the poles of the Bloch sphere represent pointer states and are stable points of attraction for any given value of $\lambda$, the dynamics defined by equation~\eqref{eq:thet1} satisfies all three requirements for a model of objective collapse. It thus serves as a minimal example of possible collapse dynamics starting from a two-state superposition. Although similar dynamics has been analysed in the context of specific objective collapse models before~\cite{Wezel_Brink}, 
it is not obvious from equation~\eqref{eq:thet1} what its physical origin could be, or how it is most naturally extended to configurations involving more than two states.

\section{Discussion}
We have shown that linear time evolution operators cannot model objective collapse. That is, linear evolution can satisfy only two out of the three minimal requirements for an objective collapse theory. It can lead to the reduction of any initial state to a single pointer state, and these pointer states can also be stable under the linear evolution. Linear time evolution cannot, however, lead to the emergence of Born's rule for the probability with which any particular pointer state is selected.

Although we only explicitly considered the possible linear flows of a two-state superposition, the result that objective collapse theories cannot be linear is general. After all, any theory for objective collapse should also be able to describe measurements involving an initial configuration superposed over two states.

That linear models cannot yield Born's rule seemingly contradicts a well-known suggested derivation of Born's rule using the concept of envariance. The resolution of this paradox lies in a hidden assumption of the suggested derivation, which relates the probabilities for ending up in a set of environmental (ancilla) states to the probability for ending up in a given system state. That this hidden assumption is not satisfied in the linear models considered here has ramifications also for various other interpretations of quantum mechanics, in which Born's rule had been suggested to emerge in essentially the same way as that suggested in the context of envariance.

The present work thus suggests that the question of how Born's rule can emerge in interpretations or modifications of quantum dynamics without axiomatically including it, remains an open problem. It also shows that a non-linear and non-unitary component is an essential ingredient for all objective collapse theories. A proof-of-principle non-linear dynamical law giving rise to Born's rule and satisfying all characteristics of quantum measurement is easily constructed for a two-state superposition. This may serve as a starting point for constructing physically realistic objective collapse models that describe both the collapse dynamics, and the way in which Born's rule emerges.

\bibliography{biblio}

\begin{thebibliography}{48}%
\makeatletter
\providecommand \@ifxundefined [1]{%
 \@ifx{#1\undefined}
}%
\providecommand \@ifnum [1]{%
 \ifnum #1\expandafter \@firstoftwo
 \else \expandafter \@secondoftwo
 \fi
}%
\providecommand \@ifx [1]{%
 \ifx #1\expandafter \@firstoftwo
 \else \expandafter \@secondoftwo
 \fi
}%
\providecommand \natexlab [1]{#1}%
\providecommand \enquote  [1]{``#1''}%
\providecommand \bibnamefont  [1]{#1}%
\providecommand \bibfnamefont [1]{#1}%
\providecommand \citenamefont [1]{#1}%
\providecommand \href@noop [0]{\@secondoftwo}%
\providecommand \href [0]{\begingroup \@sanitize@url \@href}%
\providecommand \@href[1]{\@@startlink{#1}\@@href}%
\providecommand \@@href[1]{\endgroup#1\@@endlink}%
\providecommand \@sanitize@url [0]{\catcode `\\12\catcode `\$12\catcode
  `\&12\catcode `\#12\catcode `\^12\catcode `\_12\catcode `\%12\relax}%
\providecommand \@@startlink[1]{}%
\providecommand \@@endlink[0]{}%
\providecommand \url  [0]{\begingroup\@sanitize@url \@url }%
\providecommand \@url [1]{\endgroup\@href {#1}{\urlprefix }}%
\providecommand \urlprefix  [0]{URL }%
\providecommand \Eprint [0]{\href }%
\providecommand \doibase [0]{http://dx.doi.org/}%
\providecommand \selectlanguage [0]{\@gobble}%
\providecommand \bibinfo  [0]{\@secondoftwo}%
\providecommand \bibfield  [0]{\@secondoftwo}%
\providecommand \translation [1]{[#1]}%
\providecommand \BibitemOpen [0]{}%
\providecommand \bibitemStop [0]{}%
\providecommand \bibitemNoStop [0]{.\EOS\space}%
\providecommand \EOS [0]{\spacefactor3000\relax}%
\providecommand \BibitemShut  [1]{\csname bibitem#1\endcsname}%
\let\auto@bib@innerbib\@empty
\bibitem [{\citenamefont {Peskin}\ and\ \citenamefont
  {Schroeder}(2020)}]{peskin_schroeder_2020}%
  \BibitemOpen
  \bibfield  {author} {\bibinfo {author} {\bibfnamefont {M.~E.}\ \bibnamefont
  {Peskin}}\ and\ \bibinfo {author} {\bibfnamefont {D.~V.}\ \bibnamefont
  {Schroeder}},\ }\href@noop {} {\emph {\bibinfo {title} {An introduction to
  quantum field theory}}}\ (\bibinfo  {publisher} {CRC press},\ \bibinfo {year}
  {2020})\BibitemShut {NoStop}%
\bibitem [{\citenamefont {Komar}(1962)}]{komar_1962}%
  \BibitemOpen
  \bibfield  {author} {\bibinfo {author} {\bibfnamefont {A.}~\bibnamefont
  {Komar}},\ }\href {\doibase 10.1103/physrev.126.365} {\bibfield  {journal}
  {\bibinfo  {journal} {Physical Review}\ }\textbf {\bibinfo {volume} {126}},\
  \bibinfo {pages} {365–369} (\bibinfo {year} {1962})}\BibitemShut {NoStop}%
\bibitem [{\citenamefont {Wigner}(1963)}]{wigner_1963}%
  \BibitemOpen
  \bibfield  {author} {\bibinfo {author} {\bibfnamefont {E.~P.}\ \bibnamefont
  {Wigner}},\ }\href {\doibase 10.1119/1.1969254} {\bibfield  {journal}
  {\bibinfo  {journal} {Am. J. Phys.}\ }\textbf {\bibinfo {volume} {31}},\
  \bibinfo {pages} {6–15} (\bibinfo {year} {1963})}\BibitemShut {NoStop}%
\bibitem [{\citenamefont {Adler}(2003)}]{whynotdec}%
  \BibitemOpen
  \bibfield  {author} {\bibinfo {author} {\bibfnamefont {S.~L.}\ \bibnamefont
  {Adler}},\ }\href {\doibase 10.1016/S1355-2198(02)00086-2} {\bibfield
  {journal} {\bibinfo  {journal} {Stud. Hist. Phil. Science B}\ }\textbf
  {\bibinfo {volume} {34}},\ \bibinfo {pages} {135} (\bibinfo {year}
  {2003})}\BibitemShut {NoStop}%
\bibitem [{\citenamefont {Penrose}(2014)}]{penrosegrav}%
  \BibitemOpen
  \bibfield  {author} {\bibinfo {author} {\bibfnamefont {R.}~\bibnamefont
  {Penrose}},\ }\href {\doibase 10.1007/s10701-013-9770-0} {\bibfield
  {journal} {\bibinfo  {journal} {Found. Phys.}\ }\textbf {\bibinfo {volume}
  {44}},\ \bibinfo {pages} {557} (\bibinfo {year} {2014})}\BibitemShut
  {NoStop}%
\bibitem [{\citenamefont {Everett}(1957)}]{manyworlds}%
  \BibitemOpen
  \bibfield  {author} {\bibinfo {author} {\bibfnamefont {H.}~\bibnamefont
  {Everett}},\ }\href@noop {} {\bibfield  {journal} {\bibinfo  {journal} {Rev.
  Mod. Phys.}\ }\textbf {\bibinfo {volume} {29}},\ \bibinfo {pages} {454}
  (\bibinfo {year} {1957})}\BibitemShut {NoStop}%
\bibitem [{\citenamefont {Bohm}(1952)}]{bohm}%
  \BibitemOpen
  \bibfield  {author} {\bibinfo {author} {\bibfnamefont {D.}~\bibnamefont
  {Bohm}},\ }\href@noop {} {\bibfield  {journal} {\bibinfo  {journal} {Phys.
  Rev}\ }\textbf {\bibinfo {volume} {85}},\ \bibinfo {pages} {166} (\bibinfo
  {year} {1952})}\BibitemShut {NoStop}%
\bibitem [{\citenamefont {Bohr}(1928)}]{copenhagen}%
  \BibitemOpen
  \bibfield  {author} {\bibinfo {author} {\bibfnamefont {N.}~\bibnamefont
  {Bohr}},\ }\href@noop {} {\bibfield  {journal} {\bibinfo  {journal} {Nature}\
  }\textbf {\bibinfo {volume} {121}},\ \bibinfo {pages} {580–590} (\bibinfo
  {year} {1928})}\BibitemShut {NoStop}%
\bibitem [{\citenamefont {Ghirardi}\ \emph {et~al.}(1990)\citenamefont
  {Ghirardi}, \citenamefont {Rimini},\ and\ \citenamefont {Pearle}}]{CSL}%
  \BibitemOpen
  \bibfield  {author} {\bibinfo {author} {\bibfnamefont {G.}~\bibnamefont
  {Ghirardi}}, \bibinfo {author} {\bibfnamefont {A.}~\bibnamefont {Rimini}}, \
  and\ \bibinfo {author} {\bibfnamefont {P.}~\bibnamefont {Pearle}},\
  }\href@noop {} {\bibfield  {journal} {\bibinfo  {journal} {Phys. Rev. A}\
  }\textbf {\bibinfo {volume} {42}},\ \bibinfo {pages} {78} (\bibinfo {year}
  {1990})}\BibitemShut {NoStop}%
\bibitem [{\citenamefont {Ghirardi}\ \emph {et~al.}(1986)\citenamefont
  {Ghirardi}, \citenamefont {Rimini},\ and\ \citenamefont {Weber}}]{GRW}%
  \BibitemOpen
  \bibfield  {author} {\bibinfo {author} {\bibfnamefont {G.}~\bibnamefont
  {Ghirardi}}, \bibinfo {author} {\bibfnamefont {A.}~\bibnamefont {Rimini}}, \
  and\ \bibinfo {author} {\bibfnamefont {T.}~\bibnamefont {Weber}},\
  }\href@noop {} {\bibfield  {journal} {\bibinfo  {journal} {Phys. Rev. D}\
  }\textbf {\bibinfo {volume} {34}},\ \bibinfo {pages} {580} (\bibinfo {year}
  {1986})}\BibitemShut {NoStop}%
\bibitem [{\citenamefont {Di\'osi}(1987)}]{Mastereq}%
  \BibitemOpen
  \bibfield  {author} {\bibinfo {author} {\bibfnamefont {L.}~\bibnamefont
  {Di\'osi}},\ }\href@noop {} {\bibfield  {journal} {\bibinfo  {journal} {Phys.
  Lett.}\ }\textbf {\bibinfo {volume} {120}},\ \bibinfo {pages} {377} (\bibinfo
  {year} {1987})}\BibitemShut {NoStop}%
\bibitem [{\citenamefont {Bassi}\ \emph {et~al.}(2013)\citenamefont {Bassi},
  \citenamefont {Lochan}, \citenamefont {Satin}, \citenamefont {Singh},\ and\
  \citenamefont {Ulbricht}}]{overview}%
  \BibitemOpen
  \bibfield  {author} {\bibinfo {author} {\bibfnamefont {A.}~\bibnamefont
  {Bassi}}, \bibinfo {author} {\bibfnamefont {K.}~\bibnamefont {Lochan}},
  \bibinfo {author} {\bibfnamefont {S.}~\bibnamefont {Satin}}, \bibinfo
  {author} {\bibfnamefont {T.~P.}\ \bibnamefont {Singh}}, \ and\ \bibinfo
  {author} {\bibfnamefont {H.}~\bibnamefont {Ulbricht}},\ }\href@noop {}
  {\bibfield  {journal} {\bibinfo  {journal} {Rev. of mod. phys}\ }\textbf
  {\bibinfo {volume} {85}},\ \bibinfo {pages} {471–527} (\bibinfo {year}
  {2013})}\BibitemShut {NoStop}%
\bibitem [{\citenamefont {Penrose}(1996)}]{Penrose}%
  \BibitemOpen
  \bibfield  {author} {\bibinfo {author} {\bibfnamefont {R.}~\bibnamefont
  {Penrose}},\ }\href@noop {} {\bibfield  {journal} {\bibinfo  {journal} {Gen.
  Relativ. Gravit.}\ }\textbf {\bibinfo {volume} {28}},\ \bibinfo {pages} {581}
  (\bibinfo {year} {1996})}\BibitemShut {NoStop}%
\bibitem [{\citenamefont {Van~Wezel}(2010)}]{symmetry}%
  \BibitemOpen
  \bibfield  {author} {\bibinfo {author} {\bibfnamefont {J.}~\bibnamefont
  {Van~Wezel}},\ }\href@noop {} {\bibfield  {journal} {\bibinfo  {journal}
  {Symmetry}\ }\textbf {\bibinfo {volume} {2}},\ \bibinfo {pages} {582}
  (\bibinfo {year} {2010})}\BibitemShut {NoStop}%
\bibitem [{\citenamefont {Schlosshauer}\ \emph {et~al.}(2013)\citenamefont
  {Schlosshauer}, \citenamefont {Kofler},\ and\ \citenamefont
  {Zeilinger}}]{schlosshauer_kofler_zeilinger_2013}%
  \BibitemOpen
  \bibfield  {author} {\bibinfo {author} {\bibfnamefont {M.}~\bibnamefont
  {Schlosshauer}}, \bibinfo {author} {\bibfnamefont {J.}~\bibnamefont
  {Kofler}}, \ and\ \bibinfo {author} {\bibfnamefont {A.}~\bibnamefont
  {Zeilinger}},\ }\href {\doibase 10.1016/j.shpsb.2013.04.004} {\bibfield
  {journal} {\bibinfo  {journal} {Studies in History and Philosophy of Science
  Part B: Studies in History and Philosophy of Modern Physics}\ }\textbf
  {\bibinfo {volume} {44}},\ \bibinfo {pages} {222–230} (\bibinfo {year}
  {2013})}\BibitemShut {NoStop}%
\bibitem [{\citenamefont {Marshall}\ \emph {et~al.}(2003)\citenamefont
  {Marshall}, \citenamefont {Simon}, \citenamefont {Penrose},\ and\
  \citenamefont {Bouwmeester}}]{superposmirror}%
  \BibitemOpen
  \bibfield  {author} {\bibinfo {author} {\bibfnamefont {W.}~\bibnamefont
  {Marshall}}, \bibinfo {author} {\bibfnamefont {C.}~\bibnamefont {Simon}},
  \bibinfo {author} {\bibfnamefont {R.}~\bibnamefont {Penrose}}, \ and\
  \bibinfo {author} {\bibfnamefont {D.}~\bibnamefont {Bouwmeester}},\
  }\href@noop {} {\bibfield  {journal} {\bibinfo  {journal} {Phys. Rev. Lett.}\
  }\textbf {\bibinfo {volume} {91}} (\bibinfo {year} {2003})}\BibitemShut
  {NoStop}%
\bibitem [{\citenamefont {Donadi}\ \emph {et~al.}(2020)\citenamefont {Donadi},
  \citenamefont {Piscicchia}, \citenamefont {Curceanu}, \citenamefont
  {Di\'{o}si}, \citenamefont {Laubenstein},\ and\ \citenamefont
  {Bassi}}]{underground}%
  \BibitemOpen
  \bibfield  {author} {\bibinfo {author} {\bibfnamefont {S.}~\bibnamefont
  {Donadi}}, \bibinfo {author} {\bibfnamefont {K.}~\bibnamefont {Piscicchia}},
  \bibinfo {author} {\bibfnamefont {C.}~\bibnamefont {Curceanu}}, \bibinfo
  {author} {\bibfnamefont {L.}~\bibnamefont {Di\'{o}si}}, \bibinfo {author}
  {\bibfnamefont {M.}~\bibnamefont {Laubenstein}}, \ and\ \bibinfo {author}
  {\bibfnamefont {A.}~\bibnamefont {Bassi}},\ }\href {\doibase
  10.1038/s41567-020-1008-4} {\bibfield  {journal} {\bibinfo  {journal} {Nature
  Physics}\ } (\bibinfo {year} {2020}),\ 10.1038/s41567-020-1008-4}\BibitemShut
  {NoStop}%
\bibitem [{\citenamefont {Vinante}\ \emph {et~al.}(2017)\citenamefont
  {Vinante}, \citenamefont {Mezzena}, \citenamefont {Falferi}, \citenamefont
  {Carlesso},\ and\ \citenamefont
  {Bassi}}]{vinante_mezzena_falferi_carlesso_bassi_2017}%
  \BibitemOpen
  \bibfield  {author} {\bibinfo {author} {\bibfnamefont {A.}~\bibnamefont
  {Vinante}}, \bibinfo {author} {\bibfnamefont {R.}~\bibnamefont {Mezzena}},
  \bibinfo {author} {\bibfnamefont {P.}~\bibnamefont {Falferi}}, \bibinfo
  {author} {\bibfnamefont {M.}~\bibnamefont {Carlesso}}, \ and\ \bibinfo
  {author} {\bibfnamefont {A.}~\bibnamefont {Bassi}},\ }\href {\doibase
  10.1103/physrevlett.119.110401} {\bibfield  {journal} {\bibinfo  {journal}
  {Physical Review Letters}\ }\textbf {\bibinfo {volume} {119}} (\bibinfo
  {year} {2017}),\ 10.1103/physrevlett.119.110401}\BibitemShut {NoStop}%
\bibitem [{\citenamefont {Carlesso}\ \emph {et~al.}(2016)\citenamefont
  {Carlesso}, \citenamefont {Bassi}, \citenamefont {Falferi},\ and\
  \citenamefont {Vinante}}]{carlesso_bassi_falferi_vinante_2016}%
  \BibitemOpen
  \bibfield  {author} {\bibinfo {author} {\bibfnamefont {M.}~\bibnamefont
  {Carlesso}}, \bibinfo {author} {\bibfnamefont {A.}~\bibnamefont {Bassi}},
  \bibinfo {author} {\bibfnamefont {P.}~\bibnamefont {Falferi}}, \ and\
  \bibinfo {author} {\bibfnamefont {A.}~\bibnamefont {Vinante}},\ }\href
  {\doibase 10.1103/physrevd.94.124036} {\bibfield  {journal} {\bibinfo
  {journal} {Physical Review D}\ }\textbf {\bibinfo {volume} {94}} (\bibinfo
  {year} {2016}),\ 10.1103/physrevd.94.124036}\BibitemShut {NoStop}%
\bibitem [{\citenamefont {Wit}\ \emph {et~al.}(2019)\citenamefont {Wit},
  \citenamefont {Welker}, \citenamefont {Heeck}, \citenamefont {Buters},
  \citenamefont {Eerkens}, \citenamefont {Koning}, \citenamefont {Meer},
  \citenamefont {Bouwmeester},\ and\ \citenamefont
  {Oosterkamp}}]{wit_welker_heeck_buters_eerkens_koning_meer_bouwmeester_oosterkamp_2019}%
  \BibitemOpen
  \bibfield  {author} {\bibinfo {author} {\bibfnamefont {M.~D.}\ \bibnamefont
  {Wit}}, \bibinfo {author} {\bibfnamefont {G.}~\bibnamefont {Welker}},
  \bibinfo {author} {\bibfnamefont {K.}~\bibnamefont {Heeck}}, \bibinfo
  {author} {\bibfnamefont {F.~M.}\ \bibnamefont {Buters}}, \bibinfo {author}
  {\bibfnamefont {H.~J.}\ \bibnamefont {Eerkens}}, \bibinfo {author}
  {\bibfnamefont {G.}~\bibnamefont {Koning}}, \bibinfo {author} {\bibfnamefont
  {H.~V.~D.}\ \bibnamefont {Meer}}, \bibinfo {author} {\bibfnamefont
  {D.}~\bibnamefont {Bouwmeester}}, \ and\ \bibinfo {author} {\bibfnamefont
  {T.~H.}\ \bibnamefont {Oosterkamp}},\ }\href {\doibase 10.1063/1.5066618}
  {\bibfield  {journal} {\bibinfo  {journal} {Review of Scientific
  Instruments}\ }\textbf {\bibinfo {volume} {90}},\ \bibinfo {pages} {015112}
  (\bibinfo {year} {2019})}\BibitemShut {NoStop}%
\bibitem [{\citenamefont {Arndt}\ \emph {et~al.}(1999)\citenamefont {Arndt},
  \citenamefont {Nairz}, \citenamefont {Vos-Andreae}, \citenamefont {Keller},
  \citenamefont {Zouw},\ and\ \citenamefont
  {Zeilinger}}]{arndt_nairz_vos-andreae_keller_zouw_zeilinger_1999}%
  \BibitemOpen
  \bibfield  {author} {\bibinfo {author} {\bibfnamefont {M.}~\bibnamefont
  {Arndt}}, \bibinfo {author} {\bibfnamefont {O.}~\bibnamefont {Nairz}},
  \bibinfo {author} {\bibfnamefont {J.}~\bibnamefont {Vos-Andreae}}, \bibinfo
  {author} {\bibfnamefont {C.}~\bibnamefont {Keller}}, \bibinfo {author}
  {\bibfnamefont {G.~V.~D.}\ \bibnamefont {Zouw}}, \ and\ \bibinfo {author}
  {\bibfnamefont {A.}~\bibnamefont {Zeilinger}},\ }\href {\doibase
  10.1038/44348} {\bibfield  {journal} {\bibinfo  {journal} {Nature}\ }\textbf
  {\bibinfo {volume} {401}},\ \bibinfo {pages} {680–682} (\bibinfo {year}
  {1999})}\BibitemShut {NoStop}%
\bibitem [{\citenamefont {Mooij}\ and\ \citenamefont
  {Nazarov}(2006)}]{mooij_nazarov_2006}%
  \BibitemOpen
  \bibfield  {author} {\bibinfo {author} {\bibfnamefont {J.~E.}\ \bibnamefont
  {Mooij}}\ and\ \bibinfo {author} {\bibfnamefont {Y.~V.}\ \bibnamefont
  {Nazarov}},\ }\href {\doibase 10.1038/nphys234} {\bibfield  {journal}
  {\bibinfo  {journal} {Nature Physics}\ }\textbf {\bibinfo {volume} {2}},\
  \bibinfo {pages} {169–172} (\bibinfo {year} {2006})}\BibitemShut {NoStop}%
\bibitem [{\citenamefont {Andrews}(1997)}]{andrews_1997}%
  \BibitemOpen
  \bibfield  {author} {\bibinfo {author} {\bibfnamefont {M.~R.}\ \bibnamefont
  {Andrews}},\ }\href {\doibase 10.1126/science.275.5300.637} {\bibfield
  {journal} {\bibinfo  {journal} {Science}\ }\textbf {\bibinfo {volume}
  {275}},\ \bibinfo {pages} {637–641} (\bibinfo {year} {1997})}\BibitemShut
  {NoStop}%
\bibitem [{\citenamefont {Christian}(2005)}]{christian_2005}%
  \BibitemOpen
  \bibfield  {author} {\bibinfo {author} {\bibfnamefont {J.}~\bibnamefont
  {Christian}},\ }\href {\doibase 10.1103/physrevlett.95.160403} {\bibfield
  {journal} {\bibinfo  {journal} {Physical Review Letters}\ }\textbf {\bibinfo
  {volume} {95}} (\bibinfo {year} {2005}),\
  10.1103/physrevlett.95.160403}\BibitemShut {NoStop}%
\bibitem [{\citenamefont {Squires}(1990)}]{BR}%
  \BibitemOpen
  \bibfield  {author} {\bibinfo {author} {\bibfnamefont {E.~J.}\ \bibnamefont
  {Squires}},\ }\href {\doibase 10.1016/0375-9601(90)90192-q} {\bibfield
  {journal} {\bibinfo  {journal} {Phys. Lett. A}\ }\textbf {\bibinfo {volume}
  {145}},\ \bibinfo {pages} {67–68} (\bibinfo {year} {1990})}\BibitemShut
  {NoStop}%
\bibitem [{\citenamefont {Kent}(1990)}]{MWBR}%
  \BibitemOpen
  \bibfield  {author} {\bibinfo {author} {\bibfnamefont {A.}~\bibnamefont
  {Kent}},\ }\href@noop {} {\bibfield  {journal} {\bibinfo  {journal} {Int. J.
  Mod. Phys. A}\ }\textbf {\bibinfo {volume} {5}},\ \bibinfo {pages} {1745}
  (\bibinfo {year} {1990})}\BibitemShut {NoStop}%
\bibitem [{\citenamefont {Zeh}(1999)}]{ZehBR}%
  \BibitemOpen
  \bibfield  {author} {\bibinfo {author} {\bibfnamefont {H.~D.}\ \bibnamefont
  {Zeh}},\ }\href@noop {} {\emph {\bibinfo {title} {The Physical Basis of the
  Direction of Time}}}\ (\bibinfo  {publisher} {Springer, Heidelberg},\
  \bibinfo {year} {1999})\BibitemShut {NoStop}%
\bibitem [{\citenamefont {Neumann}\ \emph {et~al.}(1955)\citenamefont
  {Neumann}, \citenamefont {Beyer}, \citenamefont {Wigner},\ and\ \citenamefont
  {Hofstadter}}]{neumann_beyer_wigner_hofstadter_1955}%
  \BibitemOpen
  \bibfield  {author} {\bibinfo {author} {\bibfnamefont {J.}~\bibnamefont
  {Neumann}}, \bibinfo {author} {\bibfnamefont {R.~T.}\ \bibnamefont {Beyer}},
  \bibinfo {author} {\bibfnamefont {E.~P.}\ \bibnamefont {Wigner}}, \ and\
  \bibinfo {author} {\bibfnamefont {R.}~\bibnamefont {Hofstadter}},\
  }\href@noop {} {\emph {\bibinfo {title} {Mathematical Foundations of quantum
  mechanics}}}\ (\bibinfo  {publisher} {Princeton University Press},\ \bibinfo
  {year} {1955})\BibitemShut {NoStop}%
\bibitem [{\citenamefont {Zurek}(1981)}]{zurek_1981}%
  \BibitemOpen
  \bibfield  {author} {\bibinfo {author} {\bibfnamefont {W.~H.}\ \bibnamefont
  {Zurek}},\ }\href {\doibase 10.1103/physrevd.24.1516} {\bibfield  {journal}
  {\bibinfo  {journal} {Physical Review D}\ }\textbf {\bibinfo {volume} {24}},\
  \bibinfo {pages} {1516–1525} (\bibinfo {year} {1981})}\BibitemShut
  {NoStop}%
\bibitem [{\citenamefont {Anglin}\ \emph {et~al.}(1997)\citenamefont {Anglin},
  \citenamefont {Paz},\ and\ \citenamefont {Zurek}}]{anglin_paz_zurek_1997}%
  \BibitemOpen
  \bibfield  {author} {\bibinfo {author} {\bibfnamefont {J.~R.}\ \bibnamefont
  {Anglin}}, \bibinfo {author} {\bibfnamefont {J.~P.}\ \bibnamefont {Paz}}, \
  and\ \bibinfo {author} {\bibfnamefont {W.~H.}\ \bibnamefont {Zurek}},\ }\href
  {\doibase 10.1103/physreva.55.4041} {\bibfield  {journal} {\bibinfo
  {journal} {Physical Review A}\ }\textbf {\bibinfo {volume} {55}},\ \bibinfo
  {pages} {4041–4053} (\bibinfo {year} {1997})}\BibitemShut {NoStop}%
\bibitem [{\citenamefont {Pearle}(2007)}]{pearle}%
  \BibitemOpen
  \bibfield  {author} {\bibinfo {author} {\bibfnamefont {P.}~\bibnamefont
  {Pearle}},\ }\href {\doibase 10.1007/bfb0104404} {\bibfield  {journal}
  {\bibinfo  {journal} {Open Systems and Measurement in Relativistic Quantum
  Theory}\ ,\ \bibinfo {pages} {195–234}} (\bibinfo {year}
  {2007})}\BibitemShut {NoStop}%
\bibitem [{\citenamefont {Di\'{o}si}(1987)}]{diosi_1987}%
  \BibitemOpen
  \bibfield  {author} {\bibinfo {author} {\bibfnamefont {L.}~\bibnamefont
  {Di\'{o}si}},\ }\href {\doibase 10.1016/0375-9601(87)90681-5} {\bibfield
  {journal} {\bibinfo  {journal} {Physics Letters A}\ }\textbf {\bibinfo
  {volume} {120}},\ \bibinfo {pages} {377–381} (\bibinfo {year}
  {1987})}\BibitemShut {NoStop}%
\bibitem [{\citenamefont {van Wezel}\ and\ \citenamefont {van~den
  Brink}(2008)}]{Wezel_Brink}%
  \BibitemOpen
  \bibfield  {author} {\bibinfo {author} {\bibfnamefont {J.}~\bibnamefont {van
  Wezel}}\ and\ \bibinfo {author} {\bibfnamefont {J.}~\bibnamefont {van~den
  Brink}},\ }\href {\doibase 10.1080/14786430802251439} {\bibfield  {journal}
  {\bibinfo  {journal} {Philosophical Magazine}\ }\textbf {\bibinfo {volume}
  {88}},\ \bibinfo {pages} {1659–1671} (\bibinfo {year} {2008})}\BibitemShut
  {NoStop}%
\bibitem [{\citenamefont {Sinha}\ \emph {et~al.}(2010)\citenamefont {Sinha},
  \citenamefont {Couteau}, \citenamefont {Jennewein}, \citenamefont
  {Laflamme},\ and\ \citenamefont
  {Weihs}}]{sinha_couteau_jennewein_laflamme_weihs_2010}%
  \BibitemOpen
  \bibfield  {author} {\bibinfo {author} {\bibfnamefont {U.}~\bibnamefont
  {Sinha}}, \bibinfo {author} {\bibfnamefont {C.}~\bibnamefont {Couteau}},
  \bibinfo {author} {\bibfnamefont {T.}~\bibnamefont {Jennewein}}, \bibinfo
  {author} {\bibfnamefont {R.}~\bibnamefont {Laflamme}}, \ and\ \bibinfo
  {author} {\bibfnamefont {G.}~\bibnamefont {Weihs}},\ }\href {\doibase
  10.1126/science.1190545} {\bibfield  {journal} {\bibinfo  {journal}
  {Science}\ }\textbf {\bibinfo {volume} {329}},\ \bibinfo {pages} {418–421}
  (\bibinfo {year} {2010})}\BibitemShut {NoStop}%
\bibitem [{\citenamefont {Pleinert}\ \emph {et~al.}(2020)\citenamefont
  {Pleinert}, \citenamefont {von Zanthier},\ and\ \citenamefont
  {Lutz}}]{pleinert_von}%
  \BibitemOpen
  \bibfield  {author} {\bibinfo {author} {\bibfnamefont {M.-O.}\ \bibnamefont
  {Pleinert}}, \bibinfo {author} {\bibfnamefont {J.}~\bibnamefont {von
  Zanthier}}, \ and\ \bibinfo {author} {\bibfnamefont {E.}~\bibnamefont
  {Lutz}},\ }\href {\doibase 10.1103/physrevresearch.2.012051} {\bibfield
  {journal} {\bibinfo  {journal} {Physical Review Research}\ }\textbf {\bibinfo
  {volume} {2}} (\bibinfo {year} {2020}),\
  10.1103/physrevresearch.2.012051}\BibitemShut {NoStop}%
\bibitem [{\citenamefont {Zurek}(2003)}]{Zurek}%
  \BibitemOpen
  \bibfield  {author} {\bibinfo {author} {\bibfnamefont {W.~H.}\ \bibnamefont
  {Zurek}},\ }\href@noop {} {\bibfield  {journal} {\bibinfo  {journal} {Phys.
  Rev. Lett.}\ }\textbf {\bibinfo {volume} {90}} (\bibinfo {year}
  {2003})}\BibitemShut {NoStop}%
\bibitem [{\citenamefont {Valentini}\ and\ \citenamefont
  {Westman}(2005{\natexlab{a}})}]{valentini_westman_2005}%
  \BibitemOpen
  \bibfield  {author} {\bibinfo {author} {\bibfnamefont {A.}~\bibnamefont
  {Valentini}}\ and\ \bibinfo {author} {\bibfnamefont {H.}~\bibnamefont
  {Westman}},\ }\href {\doibase 10.1098/rspa.2004.1394} {\bibfield  {journal}
  {\bibinfo  {journal} {Proceedings of the Royal Society A: Mathematical,
  Physical and Engineering Sciences}\ }\textbf {\bibinfo {volume} {461}},\
  \bibinfo {pages} {253–272} (\bibinfo {year}
  {2005}{\natexlab{a}})}\BibitemShut {NoStop}%
\bibitem [{\citenamefont {van Wezel}(2008)}]{vanwezelprb}%
  \BibitemOpen
  \bibfield  {author} {\bibinfo {author} {\bibfnamefont {J.}~\bibnamefont {van
  Wezel}},\ }\href {\doibase 10.1103/PhysRevB.78.054301} {\bibfield  {journal}
  {\bibinfo  {journal} {Phys. Rev. B}\ }\textbf {\bibinfo {volume} {78}},\
  \bibinfo {pages} {054301} (\bibinfo {year} {2008})}\BibitemShut {NoStop}%
\bibitem [{\citenamefont {Deutsch}(1999)}]{Deutsch_99}%
  \BibitemOpen
  \bibfield  {author} {\bibinfo {author} {\bibfnamefont {D.}~\bibnamefont
  {Deutsch}},\ }\href@noop {} {\bibfield  {journal} {\bibinfo  {journal} {Pro.
  R. Soc. London}\ }\textbf {\bibinfo {volume} {A455}},\ \bibinfo {pages}
  {3129–3137} (\bibinfo {year} {1999})}\BibitemShut {NoStop}%
\bibitem [{\citenamefont {Wallace}(2003)}]{Wallace_03}%
  \BibitemOpen
  \bibfield  {author} {\bibinfo {author} {\bibfnamefont {D.}~\bibnamefont
  {Wallace}},\ }\href@noop {} {\bibfield  {journal} {\bibinfo  {journal} {Stud.
  Hist. Phil. Mod. Phys.}\ }\textbf {\bibinfo {volume} {34}},\ \bibinfo {pages}
  {415} (\bibinfo {year} {2003})}\BibitemShut {NoStop}%
\bibitem [{\citenamefont {Wallace}(2012)}]{Wallace_12}%
  \BibitemOpen
  \bibfield  {author} {\bibinfo {author} {\bibfnamefont {D.}~\bibnamefont
  {Wallace}},\ }in\ \href@noop {} {\emph {\bibinfo {booktitle} {Many Worlds?:
  Everett, Quantum Theory, and Reality}}},\ \bibinfo {editor} {edited by\
  \bibinfo {editor} {\bibfnamefont {S.}~\bibnamefont {Saunders}}, \bibinfo
  {editor} {\bibfnamefont {J.}~\bibnamefont {Barrett}}, \bibinfo {editor}
  {\bibfnamefont {A.}~\bibnamefont {Kent}}, \ and\ \bibinfo {editor}
  {\bibfnamefont {D.}~\bibnamefont {Wallace}}}\ (\bibinfo  {publisher} {Oxford
  University Press},\ \bibinfo {year} {2012})\BibitemShut {NoStop}%
\bibitem [{\citenamefont {Valentini}\ and\ \citenamefont
  {Westman}(2005{\natexlab{b}})}]{Valentini}%
  \BibitemOpen
  \bibfield  {author} {\bibinfo {author} {\bibfnamefont {A.}~\bibnamefont
  {Valentini}}\ and\ \bibinfo {author} {\bibfnamefont {H.}~\bibnamefont
  {Westman}},\ }\href@noop {} {\bibfield  {journal} {\bibinfo  {journal} {Proc.
  R. Soc. A}\ }\textbf {\bibinfo {volume} {461}},\ \bibinfo {pages} {253}
  (\bibinfo {year} {2005}{\natexlab{b}})}\BibitemShut {NoStop}%
\bibitem [{\citenamefont {Saunders}(2004)}]{Saunders}%
  \BibitemOpen
  \bibfield  {author} {\bibinfo {author} {\bibfnamefont {S.}~\bibnamefont
  {Saunders}},\ }\href@noop {} {\bibfield  {journal} {\bibinfo  {journal}
  {Proc. R. Soc. A}\ }\textbf {\bibinfo {volume} {460}},\ \bibinfo {pages}
  {1771} (\bibinfo {year} {2004})}\BibitemShut {NoStop}%
\bibitem [{\citenamefont {Strogatz}(1998)}]{chaos}%
  \BibitemOpen
  \bibfield  {author} {\bibinfo {author} {\bibfnamefont {S.~H.}\ \bibnamefont
  {Strogatz}},\ }\href@noop {} {\emph {\bibinfo {title} {Nonlinear Dynamics and
  Chaos}}}\ (\bibinfo  {publisher} {Perseus Books},\ \bibinfo {year}
  {1998})\BibitemShut {NoStop}%
\bibitem [{\citenamefont {Gudder}(2005)}]{Fuzzy}%
  \BibitemOpen
  \bibfield  {author} {\bibinfo {author} {\bibfnamefont {S.}~\bibnamefont
  {Gudder}},\ }\href@noop {} {\bibfield  {journal} {\bibinfo  {journal}
  {Foundations of Probability and Physics}\ }\textbf {\bibinfo {volume} {3}},\
  \bibinfo {pages} {114} (\bibinfo {year} {2005})}\BibitemShut {NoStop}%
\bibitem [{\citenamefont {Redner}(2001)}]{firstpassage}%
  \BibitemOpen
  \bibfield  {author} {\bibinfo {author} {\bibfnamefont {S.}~\bibnamefont
  {Redner}},\ }\href@noop {} {\emph {\bibinfo {title} {A Guide to First-Passage
  Processes}}}\ (\bibinfo  {publisher} {Cambridge University Press},\ \bibinfo
  {year} {2001})\BibitemShut {NoStop}%
\bibitem [{\citenamefont {Mertens}(2020)}]{Lotte}%
  \BibitemOpen
  \bibfield  {author} {\bibinfo {author} {\bibfnamefont {L.}~\bibnamefont
  {Mertens}},\ }\emph {\bibinfo {title} {Spontaneous Unitary Violations and
  Effective Non-linearity in Relation to Quantum State Reduction}},\ \href@noop
  {} {Master's thesis},\ \bibinfo  {school} {University of Amsterdam} (\bibinfo
  {year} {2020})\BibitemShut {NoStop}%
\bibitem [{\citenamefont {Van~Wezel}(2007)}]{jasper}%
  \BibitemOpen
  \bibfield  {author} {\bibinfo {author} {\bibfnamefont {J.}~\bibnamefont
  {Van~Wezel}},\ }\emph {\bibinfo {title} {Quantum Mechanics and the Big
  World}},\ \href@noop {} {Ph.D. thesis},\ \bibinfo  {school} {Leiden
  University} (\bibinfo {year} {2007})\BibitemShut {NoStop}%
\end{thebibliography}%

\newpage

\appendix
\counterwithin{figure}{section}

\section{Time evolution on the Bloch sphere\label{AppA}}
To define the evolution of a general two-state superposition induced by a general time evolution generator, we consider the Bloch sphere parameterisation of the initial state:
\begin{align*}
    \ket{\psi(0)} = n e^{i\chi}\left [ e^{i\frac{\phi}{2}}\cos{(\theta/2)}\ket{0}+e^{-i\frac{\phi}{2}}\sin{(\theta/2)}\ket{1}\right].
\end{align*}
The most general form of the time evolution propagating the state forward over an infinitesimal time step $\delta t$ can be written in the form:
\begin{align*}
    \ket{\psi(\delta t)} &= e^{-i\hat{G}\delta t} \ket{\psi(0)} \notag \\
    &= \left(1 - i \hat{G} \delta t  + \mathcal{O}(\delta t^2) \right) \ket{\psi(0)}
\end{align*}
Here, $\hat{G}$ is a general $2\times 2$ matrix in the basis of the system states, which can be written in terms of eight real parameters:
\begin{align*}
    \hat{G} = \begin{pmatrix} \ket{0} & \ket{1} \end{pmatrix}
    \begin{pmatrix}
\alpha_r+i\alpha_i & \beta_r+i\beta_i\\
\gamma_r + i\gamma_i & \delta_r + i\delta_i 
\end{pmatrix}
\begin{pmatrix} \bra{0} \\ \bra{1} \end{pmatrix}.
\end{align*}
The final state $\ket{\psi(\delta t)}$ can again be parameterised on the Bloch sphere:
\begin{align*}
    \ket{\psi(\delta t)} = N e^{i{X}}& \left[ e^{i\frac{\Phi}{2}}\cos{(\Theta/2)}\ket{0} \right. \notag \\
    & \left.~~ + e^{-i\frac{\Phi}{2}}\sin{(\Theta/2)}\ket{1}\right] + \mathcal{O}(\delta t^2).
\end{align*}
From this, the time derivative of for example the parameter $\theta$ can be found exactly as $\dot{\theta}= \lim_{\delta t \to 0} (\Theta-\theta)/(\delta t)$, and similarly for the other parameters. This yields the time derivatives~\cite{Wezel_Brink,jasper}:
\begin{align}
   \dot{\theta} = & (\delta_i - \alpha_i)\sin(\theta) \nonumber \\
   & +\left((\beta_i+\gamma_i) \cos(\phi) - (\beta_r-\gamma_r) \sin(\phi) \right) \cos(\theta) \nonumber \\
   & - \left((\beta_i-\gamma_i) \cos(\phi) - (\beta_r + \gamma_r) \sin(\phi)\right), \label{eq:diff1}
\end{align}
\begin{align}
   \dot{\phi} = & (\delta_r - \alpha_r) \nonumber \\
   & - \left((\beta_r-\gamma_r) \frac{\cos(\phi)}{\sin(\theta)} + (\beta_i+\gamma_i) \frac{\sin(\phi)}{\sin(\theta)}\right) \nonumber \\
   & +\left((\beta_r+\gamma_r) \frac{\cos(\phi)}{\tan(\theta)} +(\beta_i-\gamma_i)  \frac{\sin(\phi)}{\tan(\theta)}\right), \label{eq:diff2} 
\end{align}
\begin{align}
   \dot{\chi} = & (\delta_r + \alpha_r) \nonumber \\
   & - \left((\beta_r+\gamma_r) \frac{\cos(\phi)}{\sin(\theta)} + (\beta_i-\gamma_i) \frac{\sin(\phi)}{\sin(\theta)}\right)\nonumber \\
   & + \left((\beta_r-\gamma_r) \frac{\cos(\phi)}{\tan(\theta)} +(\beta_i+\gamma_i) \frac{\sin(\phi)}{\tan(\theta)}\right), \label{eq:diff3} 
\end{align}
\begin{align}
   \frac{\dot{n}}{n} = & \frac{1}{2}(\alpha_i + \delta_i)  \nonumber \\
   & - \frac{1}{2}\left((\beta_r -\gamma_r)\sin(\phi) - (\beta_i+\gamma_i)\cos(\phi)\right)\sin(\theta)  \nonumber \\
   & + \frac{1}{2}(\alpha_i - \delta_i)\cos(\theta)
   . \label{eq:diff4}
\end{align}
From these equations it is clear that the change in the parameters $\phi$ and $\theta$ only depends on the instantaneous values of $\phi$ and $\theta$ themselves, and not on the overall phase $\chi$ or normalisation $n$. These are therefore gauge degrees of freedom and can be arbitrarily normalised to $n=1$ and $\chi=0$ at any moment in time without affecting any observable degrees of freedom.

\section{Born's rule from envariance\label{AppB}}
For completeness, we will reproduce the central steps in the suggested derivation of Born's rule in ref.~\onlinecite{Zurek}, in terms of the current formalism and notation.

The principle idea of envariance is that the statistics of local measurement outcomes on a quantum state cannot be influenced by any operation on a different, causally disconnected system. If this condition were violated, instantaneous communication between the two systems would be possible, violating the assumption of them being causally disconnected. The implications of this observation become clear when considering an entangled state of the form:
\begin{align}
    \ket{\psi} = \alpha \ket{0}\ket{a} + \beta \ket{1}\ket{b}.
\end{align}
Here, the states $\ket{0}$ and $\ket{1}$ denote the local system states, while $\ket{a}$ and $\ket{b}$ are states of the causally disconnected environment. Because the phases of $\alpha$ and $\beta$ can be altered by local unitary operations on the environmental states, they cannot influence the probabilities for finding $\ket{0}$ and $\ket{1}$ in a local measurement on the system. Assuming that the measurements are unbiassed, in the sense that they do not \emph{a priori} favour one of the system states, the probabilities can then depend only on the magnitudes of the weights in the state to be measured~\cite{Zurek}.

Next, consider the equal-weight entangled state:
\begin{align}
    \ket{\psi} = \alpha \left( \ket{0}\ket{a} + \ket{1}\ket{b} \right).
    \label{eq:env2}
\end{align}
The local, unitary swap operation on the system is defined as $\hat{U}_{\text{s}} = \ket{0}\bra{1} + \ket{1}\bra{0}$, and similarly we can define a unitary swap operation that acts locally on the environment as $\hat{U}_{\text{e}} = \ket{a}\bra{b} + \ket{b}\bra{a}$. The state of equation~\eqref{eq:env2} has the invariant property that a local swap on the system can be undone by a local swap on the environment, so that $\hat{U}_{\text{e}} \hat{U}_{\text{s}} \ket{\psi} = \ket{\psi}$. In other words, the effect of a local operation on the environment is equivalent to the effect of a local operation on the system: $\hat{U}_{\text{s}} \ket{\psi} = \hat{U}_{\text{e}}^{-1} \ket{\psi} = \hat{U}_{\text{e}}\ket{\psi}$. The swap operation on a causally disconnected environment cannot influence the statistics of local measurement outcomes on the system. But again assuming that the measurement is unbiassed, the outcome statistics when measuring the state $\hat{U}_{\text{e}}\ket{\psi}$ cannot be different from the statistics when measuring $\hat{U}_{\text{s}}\ket{\psi}$. Swapping $\ket{0}$ and $\ket{1}$ thus has no effect on the respective probabilities for registering these states, and hence their probabilities must be equal.

Notice that these arguments do not require the environmental states to actually exist or be present. In fact, since local actions on the environment cannot influence the statistics of local measurement outcomes on the system, we could consider an extreme case in which the environmental degree of freedom is destroyed (without measuring it) before the system is measured. Since the destruction of the environment cannot influence the statistics observed of the system, the probabilities for registering any particular outcome must be independent of whether or not the environment actually exists.

Extending the argument that equal weights yield equal probabilities, we can consider an entangled state involving arbitrarily many system and environmental states:
\begin{align} 
\label{eq: initial}
    \ket{\psi} = \sum_{k=1}^N \alpha_k \ket{k}\ket{e_k}.
\end{align}
Here, $\ket{k}$ signify system states, while $\ket{e_k}$ denote states of the environment. If the weights $\alpha_k$ are equal for any pair of labels $k'$ and $k''$, then the state $\ket{\psi}$ is left invariant by the consecutive swaps $\hat{U}_{\text{e}} = \ket{e_{k'}}\bra{e_{k''}} + \ket{e_{k''}}\bra{e_{k'}}$ and $\hat{U}_{\text{s}} = \ket{{k'}}\bra{{k''}} + \ket{{k''}}\bra{{k'}}$. Using the same argument as before, we then conclude that any subset of states with equal weights within a larger superposition must all have equal probabilities of being registered in a local measurement on the system.

The final step in the proposed derivation of Born's rule then concerns a superposition with unequal weights:
\begin{align}
    \ket{\psi} = \alpha \ket{0}\ket{a} + \beta \ket{1}\ket{b}.
\end{align}
Because the arguments based on envariance are based on the possible existence of environmental states, and do not require the environment to really exist or be present, we may assume the environmental Hilbert space to be arbitrarily large. It is then always possible to identify a basis for the environmental states in which the full state can be expressed as an equal weight superposition:
\begin{align}
    \ket{\psi} = \sqrt{\frac{1}{N}} \left[ \sum_{i=1}^n \ket{0}\ket{i} + \sum_{j=n+1}^{n+m} \ket{1}\ket{j} \right].
    \label{eq:zurekstateapp}
\end{align}
Here, the rational fractions $n/N$ and $m/N$ can be made to approximate the real numbers $\alpha^2$ and $\beta^2$ with arbitrary precision~\cite{Zurek}. Because the weights of all components in this state are equal, we would expect equal probabilities for registering any of them. The precise meaning of this, however, becomes clear only when we explicitly consider the swap operations whose product leaves the state invariant. In particular, the swap of system states, $\hat{U}_{\text{s}} = \ket{{0}}\bra{{1}} + \ket{{1}}\bra{{0}}$, cannot be undone by a swap operation on the environment. The only exception is the special case $m=n$, which would imply we had an equal-weight superposition with $\alpha=\beta$ to begin with. To find a combination of operations that does leave the state invariant, we need to consider the possible existence of a second environment, which we may assume to be causally disconnected from both the system and the first environment:
\begin{align}
    \ket{\psi} = \sqrt{\frac{1}{N}} \left[ \sum_{i=1}^n \ket{0}\ket{i}\ket{e_i} + \sum_{j=n+1}^{n+m} \ket{1}\ket{j}\ket{e_j} \right].
    \label{eq:zurekstateapp2}
\end{align}
In this state, a combined swap on the system and the first environment, $\hat{U}_{\text{s}} = \ket{{0}}\ket{i}\bra{{j}}\bra{1} + \ket{{1}}\ket{j}\bra{i}\bra{{0}}$, can be undone by a local swap on the second environment, $\hat{U}_{\text{e}} = \ket{{e_i}}\bra{{e_j}} + \ket{{e_j}}\bra{{e_i}}$. Because the local swap on the second environment cannot influence the outcome statistics of any `local' measurements of the system and the first environment, the probabilities of registering any of the states $\ket{0}\ket{i}$ or $\ket{1}\ket{j}$ must all be equal, and equal to $1/N$.

Notice that in this case, the first environment must actually be present. Combined swap operations on both the system and the first environment can be undone by swaps on the second environment, but swap operations on the system alone cannot. Because of this subtlety, it is \emph{not} generally true that the probability for a local measurement on the system to register $\ket{0}$ is equal to the sum of probabilities for any of the $n$ states $\ket{0}\ket{i}$ to be registered. That is, the probability to register $\ket{0}$ can \emph{not} be concluded to be $n/N$. There is an essential difference between on the one hand a local measurement on the system alone registering $\ket{0}$, and on the other hand a combined measurement of the system and the first environment registering any of the states $\ket{0}\ket{i}$. In the first case, the measurement process does not involve the first environment and the density matrix for the system and first environment after the measurement will be:
\begin{align*}
    \rho = \frac{1}{N}\left( \sum_{i,i'=1}^n \ket{0}\ket{i}\bra{i'}\bra{0} + \sum_{j,j'=n}^{n+m} \ket{1}\ket{j}\bra{j'}\bra{1} \right).
\end{align*}
In the second case, the measurement process must register one of the states $\ket{0}\ket{i}$ (on which the swap $\hat{U}_{\text{s}}$ operates) and the density matrix thus becomes:
\begin{align}
    \rho = \frac{1}{N}\left( \sum_{i=1}^n \ket{0}\ket{i}\bra{i}\bra{0} + \sum_{j=n}^{n+m} \ket{1}\ket{j}\bra{j}\bra{1} \right).
\end{align}
Even though these two matrices become the same if averaged over the environmental states, they are fundamentally different, and equal probabilities in one do not imply equal probabilities in the other. This explains how linear objective collapse models can yield statistics that are inconsistent with Born's rule, even though all steps of the envariance based argumentation do apply.

\end{document}